\documentclass[11pt]{article}
\pdfoutput=1
\usepackage{jheppub}
\usepackage{graphicx}
\usepackage{dcolumn}
\usepackage{bm}
\usepackage{amsmath,amssymb,amscd}
\usepackage{longtable}
 \usepackage{slashed}

\newcommand{\be}{\begin{eqnarray}}
\newcommand{\ee}{\end{eqnarray}}
\newcommand{\nn}{\nonumber}
\newcommand{\bn}{\begin{enumerate}}
\newcommand{\en}{\end{enumerate}}

\newcommand{\beq}{\begin{equation}}
\newcommand{\eeq}{\end{equation}}
\newcommand{\bea}{\begin{equation}\begin{aligned}}
\newcommand{\eea}{\end{aligned}\end{equation}}

\def\CI{{\cal I}}

\def\CN{{\cal N}}

\def\a{\alpha}
\def\b{\beta}
\def\g{\gamma}
\def\d{\delta}

\def\s{\sigma}

\def\t{\tau}

\def\dd{{\rm d}}

\def\goto{\rightarrow}

\def\p{\partial}

\def\det{{\rm det}}

\DeclareMathOperator{\Tr}{Tr}

\newcommand{\ket}[1]{\vert  #1\rangle}

\newcommand{\wbar}[1]{{\overline{#1}}}
\newcommand{\wtilde}[1]{\widetilde{#1}}

\title{2D Seiberg-like dualities with an  adjoint matter}

\author[a]{Kyoungho Cho,}
\author[b,c]{Hyungchul Kim,}
\author[d]{Jaemo Park}

\affiliation[a]{
Department of Physics, Sogang University, \\
 Mapo-gu, Seoul 121-742, Korea
}
\affiliation[b]{
Center of Mathematical Sciences and Applications, Harvard University, \\
Cambridge, 02138, USA
}
\affiliation[c]{
Jefferson Physical Laboratory, Harvard University, \\
Cambridge, MA 02138, USA
}
\affiliation[d]{
Department of Physics, POSTECH, \\
Pohang 790-784, Korea
}

\emailAdd{khcho23@sogang.ac.kr}
\emailAdd{hyungchul\_kim@g.harvard.edu}
\emailAdd{jaemo@postech.ac.kr}

\abstract{
We consider the analogue of Kutasov-Schwimmer-Seiberg duality for two-dimensional $\CN=(2,2)$ $U(k)$ gauge theory with one adjoint $X$ with the superpotential $\Tr X^{l+1}$ and with fundamental and anti-fundamental chiral multiplets. We give the evidences for the proposed dualities
by analytically proving that the elliptic genus of dual pair coincides with each other. For some of the dual pairs flowing to the superconformal field theory, we show the nonperturbative truncation of the chiral ring. For the theory with one adjoint and $N_f >k$ fundamental fields,
we argue the theory exhibits mass gap.  }

\begin{document}

\maketitle

\section{Introduction}
Recently there has been much progress on the understanding the Seiberg-like dualities in lower dimensions than 4.
With the help of the recently developed localization results, substantial evidences were cumulated, especially in 3-dimensions.
There's a close relation between the dualities in 4-dimensions and 3-dimensions \cite{Aharony:2013dha}, and we also expect the similar relation holds between 3-dimensions and 2-dimensions.
Indeed 2-dimensional Seiberg-like dualities  for $\CN= (2,2)$ $U(k)$ gauge theory with fundamental chiral multiplets  with/without anti-fundamental chiral multiplets
 were studied in \cite{ Benini:2012ui, Benini:2013xpa, Benini:2014mia, Gadde:2013ftv, Gomis:2014eya, Gadde:2015wta}, and the elliptic genus was computed to give the evidences for such dualities. The peculiar feature is that such duality holds for asymptotically free theories as well, while in higher dimensions
 the duality holds for superconformal field theories (SCFT). Other aspects of 2-dimensional dualities were explored in
 \cite{Horinon, Jockers, Sharpe, Orlando}.
 In 3 and 4 dimensions, in addition to  fundamental/anti-fundamental matter fields, one can also consider dualities with 2nd rank tensor matter field, so-called Kutasov-Schwimmer-Seiberg dualities and their 3-dimensional analogues\cite{Kutasov:1995np, Kutasov:1995ss, Niarchos:2008jb, Kapustin:2011vz, Kim:2013cma, Hwang:2015wna}.
 In fact, there's conjecture about such dualities for 2-dimensional theories in \cite{Gomis:2014eya} in the context of AGT correspondence. The authors of \cite{Gomis:2014eya}  gives the evidences for such dualities by checking the $S^2$ partition function.
 Here we consider the $U(k)$ gauge theory with one adjoint and with fundamental /(anti-fundamental) chiral fields and show analytically that the elliptic genus
 of the dual pair coincides with each other, thereby providing additional evidences. For theories with adjoint and fundamental matter fields, the theory exhibits the mass gap, which is similar
 to $U(k)$ gauge theory with fundamental chiral multiplets, which leads to the non-linear $\sigma$-model with the target space Grassmannian.
 In fact, the argument for mass gap for the theory with adjoint and fundamental fields is similar to Grassmannian model \cite{Witten:1993xi}.
 For theories  with adjoint and the same number of fundamentals/anti-fundamentals, the theory flows to SCFT.
 In this case, we also work out the chiral ring elements. In 3-dimensions and 4-dimensions,  nonperturbative truncation of the
 chiral ring occurs and we also check that this also occurs in our 2-dimensional cases as well.

 The contents of the paper are as follows.
 In the section 1, we introduce the basics of elliptic genera. We mainly work with the elliptic genus with the Ramond-Ramond boundary conditions.
 However, to work out the chiral ring elements, we have to use the elliptic genus with NS-NS boundary conditions.
 The relation between the R and NS sector is standard, known as spectral flow, which we summarize.
 In the section 2, we work out the elliptic genus and show that the dual pairs have the same elliptic genus.
 For $U(k)$ theories with an adjoint and fundamentals, we check it exhibits the mass gap numerically.
 For $U(k)$ theories with an adjoint and the same number of fundamentals/anti-fundamentals, the theory flows to SCFT
 and we compare their chiral ring structures as well. As occurring in higher dimensions, nonperturbative truncation of the chiral
 ring elements is observed.
 For $U(k)$ theories with different number of fundamental and anti-fundamental matter fields, we find that the theories have both discrete vacua and noncompact Higgs branches.
 This duality can be also obtained from the theory with the same number of fundamentals and anti-fundamentals and give the mass
 to anti-fundamentals.
 In the section 3, we obtain the 2-dimensional dualities from 3-dimensional dualities via dimensional reduction on a circle. The 3-dimensional dualities, in turn, can be obtained from the 4-dimensional dualities\cite{Nii:2014jsa}. As explained in \cite{Aharony:2016jki}, in order to to have dualities of $U(N)$ gauge group in 2-dimensions, non-zero FI-term should be turned on, which we assume in the subsequent discussions.
 In the section 4, we argue the existence of the mass gap for theories with an adjoint and fundamentals.
 The theory without adjoint leads to the Grassmannian model, and its low energy theory is described by gauged $U(k)/U(k)$ WZW
 model. It would be interesting to work out the analogue for the theory with an adjoint.
 In the appendix, we provide the explicit expressions for $\chi_y$ genus and Witten index for the theory with an adjoint and fundamentals.

\section{Elliptic genus of theories}
In this section, we  review the basic facts about the elliptic genus and chiral primaries in 2d SCFTs.
The elliptic genus \cite{Witten:1993jg, Benini:2013nda, Benini:2013xpa} is computed in the RR sector,
\begin{align}
Z(q,y)=\Tr_{\textrm{RR}} (-1)^F q^{H_L}y^{J_L}
\end{align}
where $H_L$ is the left-moving Hamiltonian and $J_L$ is the left-moving $U(1)$ R-charge. When the theory flows to a superconformal field theory (SCFT) in IR, $H_L$ and $J_L$ are identified as zero mode generators $L_0, J_0$ of the $N=2$ superconformal algebra respectively. When the theory does not flow to SCFT, $J_L$ takes discrete values.
In later sections, we are interested in the chiral ring structures of various theories. For this purpose, one had better look for the elliptic genus defined in the NSNS sector \cite{Gadde:2013ftv},
\begin{align}
\CI(q,y)=\Tr_{\textrm{NSNS}} (-1)^F q^{L_0}y^{J_0}~.
\end{align}

The R sector and the NS sector are connected by continuously changing the boundary conditions for fermions. This is known as the spectral flow \cite{Schwimmer:1986mf, Lerche:1989uy}. The relation of $N=2$ superconformal algebras between the R sector and the NS sector is given by
\begin{align}
&j_{\textrm{R}}=j_{\textrm{NS}}-\frac{c}{6}~,
\\
&h_{\textrm{R}}=h_{\textrm{NS}} - \frac{1}{2}j_\textrm{NS} + \frac{c}{24}~.
\end{align}
Thus the elliptic genus $Z(q,y)$ and the superconformal index $\CI(q,y)$ are related by
\begin{align}
&Z(q,y)=-a y^{-\frac{c}{6}}\CI(q,q^{-\frac{1}{2}}y)
\end{align}
where $j$ is the left-moving $U(1)$ R-charge and $h$ is the left moving conformal dimensions.\footnote{The additional factor $-a$ occurs
due to the definition of $(-1)^F$ in NS and R sectors and the regularization of the path integral of the elliptic genus.}


The anti-commutation relation between the two supersymmetry generators in the $N=2$ superconformal algebra is
\begin{align}
\{G^-_{r},G^+_{s}\}=2L_{r+s}-(r-s)J_{r+s}+(c/3)(r^2-1/4)\delta_{r+s,0}
\end{align}
where $r,s$ run over half-integral values in the NS sector and over integral values in the R sector. Chiral and anti-chiral states are states in the NS sector satisfying
\begin{align}
&G^+_{-1/2}{\ket \phi}=0 \qquad \textrm{chiral,} \label{chiral}\\
&G^-_{-1/2}{\ket \phi}=0 \qquad \textrm{anti-chiral.}\label{antichiral}
\end{align}
 Chiral (resp. anti-chiral) primary states satisfy, in addition to \eqref{chiral} (resp. \eqref{antichiral})
\begin{align}
G^-_{n+1/2}{\ket \phi}=G^+_{n+1/2}{\ket \phi}=0 \quad \textrm{for $n\geq 0$.}\
\end{align}
In a unitary theory, $G^+_{-1/2}=(G^-_{1/2})^\dagger$. The chiral primary states have the dimension $h$ and left-moving $U(1)$ charge $j$ satisfying $h=j/2$. Similarly, the anti-chiral primary states have $h=-j/2$. The operator algebra of  chiral (anti-chiral) primary fields forms a ring \cite{Lerche:1989uy}. Under the spectral flow, the chiral primary states flow to the ground states of the Ramond sector which satisfy
\begin{align}
\{G^-_0,G^+_0\}{\ket \phi}=0~.
\end{align}
The ground states in the Ramond sector satisfy $h=c/24$.

The contributions of chiral primary states to the superconformal index can be obtained by a deformation $q\rightarrow qt$ and $y\rightarrow yt^{-1/2}$ and a limit,
\begin{align}
\lim_{t\rightarrow 0}\CI(qt,yt^{-1/2})=\lim_{t\rightarrow 0}\Tr_{\textrm{NSNS}}(-1)^F(qt)^{L_0}(yt^{-1/2})^{J_0}.
\end{align}
Only states satisfying $h=j/2$ survive in the limit.

For example, the superconformal index for a free chiral superfield $\Phi$ is given by
\begin{align}
\CI_{\Phi}(q,y,a)& = \triangle(q,y,a)  = \prod_{i=0}^{\infty} \frac{(1-ay^{-1}q^{i+\frac{1}{2}})(1-a^{-1}y q^{i+\frac{1}{2}})}{(1-aq^i)(1-a^{-1}q^{i+1})}
 \end{align}
where we assume chiral superfield has a left-moving $U(1)$ R-charge $0$.  Instead when a chiral superfield have a left-moving $U(1)$ R-charge $r$ then the superconformal index can be obtained by
\begin{align}
\CI_{\Phi,r}(q,y,a)=\triangle(q,y,a(yq^{1/2})^r)
\end{align}
Contributions from the chiral primary states can be obtained by
\begin{align}
\lim_{t\rightarrow 0}\CI_{\Phi,r}(qt,yt^{-1/2},a)
&=\lim_{t\rightarrow 0}\triangle(qt,yt^{-1/2},a(yq^{1/2})^r)\nonumber\\
&=\lim_{t\rightarrow 0} \prod_{i=0}^{\infty} \frac{(1-a(yq^{1/2})^ry^{-1}q^{i+\frac{1}{2}}t^{i+1})(1-a^{-1}(yq^{1/2})^{-r}y q^{i+\frac{1}{2}}t^{i})}{(1-a(yq^{1/2})^rq^it^i)(1-a^{-1}(yq^{1/2})^{-r}q^{i+1}t^{i+1})} \nonumber\\
&=\frac{1-a^{-1}y^{1-r}q^{(1-r)/2}}{1-ay^rq^{r/2}}~.
\end{align}
The denominator and the numerator are the contributions of a scalar $\phi$ and a left-moving fermion $\wbar\psi_-$ of the chiral multiplet $\Phi$ respectively.
Because the chiral primary states of the NS sector flows to the ground states of the Ramond sector the contributions of chiral primary states can be also seen from the elliptic genus.

\subsection{$U(1)$ gauge theory with $N_f$ chiral multiplets}
Let's consider a simple gauge theory,  $U(1)$ gauge theory with $N_f$ chiral multiplets of charge 1, $CP^{N_f-1}$ model.
For general gauge theories, the elliptic genus is worked out by evaluating Jeffrey-Kirwan (JK) residues.
Elliptic genus of the theory is given by \cite{Benini:2013nda}
\begin{align}
Z(\t,z,\xi_\a)&=\sum_{u_i\in \mathfrak M^{+}_{\textrm{sing}}} \oint _{u=u_i}\dd u \frac{i\eta(q)^3}{\theta_1(q,y)} \prod_{k=1}^{N_f}\frac{\theta_1(\tau|u-\xi_k-z)}{\theta_1(\tau|u-\xi_k)}\nonumber\\
&=\sum_{\a=1}^{N_f}\;\prod_{\b \neq \a}\frac{\theta_1(\tau|-z+\xi_\a-\xi_\b)}{\theta_1(\tau|\xi_\a-\xi_\b)}
\end{align}
where $q=e^{2\pi i \tau}$, $y=e^{2\pi i z}$ and $\tau$ is the complex structure of a torus and $z$ is a holonomy for the left-moving $U(1)$ R-symmetry and $\xi_\a$, $\a=1,\ldots,N_f$ are holonomies for the $SU(N_f)$ flavor symmetry with a constraint $\sum_\alpha \xi_\alpha = 0$. $\mathfrak M^{+}_{\textrm{sing}}$ consists of $N_f$ simple poles at $u_i=\xi_i$ associated with positive charges.

Single-valuedness condition requires $y^{N_f}=1$ so the elliptic genus reduces to
\begin{align}\label{chi_y_CPn}
Z(\tau,z,\xi_\a)|_{y^{N_f}=1}=y^{-(N_f-1)/2}(1+y+\cdots+y^{N_f-1})
\end{align}
This has been checked numerically.
The elliptic genus gets contributions only from ground states because the theory develops mass gap in IR.
The UV interpretation is that we have nonlinear $\sigma$ model whose target space is $CP^{N_f-1}$ and each
factor of $y$ represents the cohomology ring elements of $CP^{N_f-1}$ \cite{Witten:elliptic}.
The truncation of the elliptic genus implies the cohomology ring relation $y^{N_f}=c$ for a suitable c number.
Due to the single-valuedness of the elliptic genus we set $y^{N_f}=1$.

\subsection{$U(k)$ with $N_f$ fundamentals and one adjoint}
In this subsection, we consider a 2d dual pair and show that the elliptic genus coincides with each other.
One theory is $U(k)$ gauge theory with $N_f$ fundamental chiral multiplets and one adjoint chiral multiplets  $X$ with
the superpotential
\begin{equation}
W=\Tr X^{l+1}
\end{equation}
The symmetries and charges of the theory are
\begin{equation}
\begin{array}{c|ccc}
& U(k) & SU(N_f) & U(1)_L \\
\hline
Q & \square  & \overline \square& 0 \\
X & \bold{Ad} & \bold{1} & \frac{1}{l+1}
\end{array}
\end{equation}
where $U(1)_L$ is the left-moving $U(1)$ R-symmetry.

One-loop determinant is
\begin{multline}\label{EG_1loop_NfAd}
Z_\text{1-loop} = (-1)^k \frac1{k!} \bigg( \frac{2\pi \eta(q)^3}{\theta_1(q,y^{-1})} \bigg)^k \bigg( \prod_{i\neq j}^k \frac{\theta_1(\tau|u_i - u_j)}{\theta_1(\tau|u_i - u_j - z)} \bigg) \times \, \\
 \bigg( \prod_{i,j=1}^k \frac{\theta_1(\tau|u_i - u_j + \frac{1}{l+1}z - z)}{\theta_1(\tau|u_i - u_j + \frac{1}{l+1}z)} \bigg)
 \bigg(\prod_{i=1}^k \prod_{\alpha=1}^{N_f} \frac{\theta_1(\tau| u_i - \xi_\alpha - z)}{\theta_1(\tau|u_i - \xi_\alpha)}\bigg) \, \dd^ku \;.
\end{multline}

We have introduced gauge holonomies $u_i$, $i=1,\ldots,k$ and flavor holonomies $\xi_\alpha$, $\a=1,\ldots,N_f$ for $SU(N_f)$ symmetry with $\sum \xi_\alpha = 0$. The first line of \eqref{EG_1loop_NfAd} comes from the vector multiplet.
In the second line, the first fraction comes from the adjoint chiral multiplet and the second fraction comes from the fundamental chiral multiplets. The adjoint chiral multiplet has a left-moving $U(1)$ R-charge $\frac{1}{l+1}$ fixed by the superpotential $W=\Tr X^{l+1}$.

We have fixed the sign of the one-loop determinant by $(-1)^k$. A sign of an elliptic genus depends on the number of decoupled massive chiral fields, which can be seen from $Z_{\Phi, r=1/2}=-1$ where the left-moving R-charge $r$ is fixed by a superpotential $W=\Phi^2$. But the number of decoupled massive chiral fields can be arbitrary so the sign is ambiguous. It is reasonable to fix the sign to have positive Witten index or consistent Renormalization Group (RG)flows to the known theories.
With the sign $(-1)^k$, as we will see, the Witten index obtained by a limit $z\rightarrow 0$ is a positive integer. The sign is consistent with that of the theory without the adjoint chiral field.  When $l=1$ the adjoint field become massive and its contribution in \eqref{EG_1loop_NfAd} becomes $(-1)^{k^2}$ which can be seen by using $\theta_1(\tau|-a)=-\theta_1(\tau|a)$. Thus the elliptic genus becomes that of the theory without the adjoint field if we have a sign factor $(-1)^{k+k^2}=1$. Furthermore, it is also consistent with the dual description of the theory as we will see.

The holonomies $u_i$ take values in $T^2$ and we have
\begin{align}
Z_{\textrm{1-loop}}(\tau, z, u_1+a+b\tau,u_2\ldots,u_k)=y^{bN_f}Z_{\textrm{1-loop}}(\tau, z, u_1,u_2,\ldots,u_k)
\end{align}
for $a,b\in \mathbb Z$. Single-valuedness of the one-loop determinant requires $y^{N_f}=1$, i.e. $z\in \mathbb Z / N_f$. It reflects the fact that the left-moving R-symmetry $U(1)_L$ of the theory is anomalous, so that  $U(1)_L$ is broken to $ Z_{N_f}$.

JK residue is evaluated by \cite{Benini:2013xpa}.
\begin{align}
 Z(\tau,z,\xi_\a) &=\frac{1}{(2\pi i)^k} \sum_{u^*\in \mathfrak M^*_{\textrm{sing}}} \oint_{u=u^*}  Z_{\textrm{1-loop}}(\tau,z,u,\xi_\a)
\end{align}
where $\mathfrak M^*_{\textrm{sing}}$ can be chosen to poles associated with positive charges so it is the set of solutions of poles
\begin{align}
&u_i=\xi_\a \label{Pole_F}\\
&u_i=u_j-\frac{1}{l+1}z \label{Pole_A}~.
\end{align}
Note that the poles $u_i=u_j+z$ from the gauge sector do not have contributions. Suppose a pole $u_{\bar i}=u_{\bar j}+z$ is picked up together with a pole at $u_{\bar j}=\xi_{\bar \a}$. Then the numerator $u_{\bar i}-\xi_{\bar \a}-z$ vanishes. That is the reason why the poles from gauge sector do not have contributions.
The two types of poles have the charge covector $(1,0)$ and $(1,-1)$ respectively in an $i$-$j$ plane.
The JK residue gets contributions only from linearly independent charge covectors. Thus pole configurations like $\{u_1=u_2-\frac{1}{l+1}z, u_2=u_3-\frac{1}{l+1}z, u_3=u_1-\frac{1}{l+1}z, \cdots \}$  do not contribute to the JK residue. If all poles are chosen from \eqref{Pole_A} the charge covectors are linearly dependent so at least one pole should be chosen from \eqref{Pole_F}.
Furthermore, if any two of poles are chosen to be the same as in $\{ u_1=\xi_1$, $u_2=\xi_1, \cdots\}$, we get a zero from the numerator of the gauge sector, i.e. $u_i-u_j=0$.
Thus one can parametrize contributing poles by ordered sequences $\vec{n}=(n_1,n_2,\ldots,n_{N_f})$ with $n_\a\geq 0$ and $\sum_\a n_a=k$. If $n_\b=0$ it means that a pole $u_i=\xi_\b$ is not chosen. If $n_\b$ is non-zero it corresponds to a case that $n_\b$ poles have a form of
\begin{align}
u^*_{i_1}=\xi_\b, u^*_{i_2}=u^*_{i_1}-\frac{1}{l+1}z, u^*_{i_3}=u^*_{i_2}-\frac{1}{l+1}z, \cdots, u^*_{i_{n_\b}}=u^*_{i_{n_\b-1}} -\frac{1}{l+1}z
\end{align}
where indices ${i_1}, \ldots, {i_{n_\b}}$ are in $\{1, \ldots, k\}$ and distinct. It can be written as
\begin{align}
u^*_{i_{m_\b+1}}=\xi_\b-m_\b\frac{1}{l+1}z \qquad \textrm{for $m_\b=0,\ldots, n_\b-1$} \label{Pole_n}
\end{align}
The total number of poles is $k$ so we have $\sum_\a n_\a=k$. Permutations of $k$ poles lead to the same residue so it cancels the Weyl group dimension $|W|=k!$. The form of poles \eqref{Pole_n} and a replacement $\prod_{i=1}^{k}\rightarrow \prod_{\a=1}^{N_f}\prod_{m_\a=0}^{n_\a-1}$ lead to
\begin{align}
 Z(\tau,z,\xi_\a) &=\frac{1}{(2\pi i)^k} \sum_{\vec{n} \phantom{,} \textrm{s.t}\, |\vec{n}|=k} \oint_{u_1=u^*_{1}} \cdots \oint_{u_k=u^*_{k}} Z_{\textrm{1-loop}}(\tau,z,u,\xi_\a)   \nonumber\\
&= (-1)^k \sum_{\vec{n} \phantom{,} \textrm{s.t}\, |\vec{n}|=k}
  \prod_{\a,\b=1}^{N_f} \prod_{m_\a =0}^{n_\a-1} \frac{\theta_1(\tau|\xi_\a -\xi_\b +(n_\b -m_\a-l-1)\frac{z}{l+1})}{\theta_1(\tau|\xi_\a -\xi_\b +(n_\b -m_\a)\frac{z}{l+1})}  ~.\label{EG_U(k)NfAd}
\end{align}
after many cancellations between the gauge sector and matter sector.

The elliptic genus do not get any contribution from configurations of $\{n_1,...,n_N\}$ which contains $n_\a$ such that $n_\a > l$. It can be seen from the fact that the numerator of \eqref{EG_U(k)NfAd} is zero if $\a=\b$ and $m_\a=n_\a-l-1$ which can be satisfied for $n_\a>l$ because $0\leq n_\a-l-1\leq n_\a-1$.
Thus one can consider only the $\{n_1,...,n_N\}$ configurations where $n_\a$ are restricted by $\sum_\a n_\a=k$ and $0\leq n_\a \leq l$.

One can also see that all contributions from the fundamental fields are canceled out and only vacuum contributions survive.
We have computed the elliptic genus explicitly and expanded it in powers of $q=e^{2i\pi \t}$. We have checked numerically that all higher order terms of $q$ become zero by the single-valuedness condition $y^{N_f}=1$
so the elliptic genus gets contributions only from $q^0$ terms. It implies that the theory is massive because the $q^0$ terms correspond to the ground states of the theory.

Let us compute the Witten index of the theory. The elliptic genus is reduced to the Witten index in the limit $z\goto 0$, i.e. $y\rightarrow 1$. In the limit $z\goto 0$, the factors of the form, $\frac{\theta_1(\tau|\xi_\a -\xi_\b +(n_\b -m_\a-l-1)\frac{z}{l+1})}{\theta_1(\tau|\xi_\a -\xi_\b +(n_\b -m_\a)\frac{z}{l+1})}$ becomes 1 if $\a\neq \b$. Thus non-trivial contributions arise from when $\a=\b$.
\begin{align}
 \lim_{z\goto 0}Z^A(\t,z,\xi)
& = (-1)^k \lim_{z\goto 0}\sum_{\vec{n} \phantom{,} \textrm{s.t}\, |\vec{n}|=k}
  \prod_{\a=1}^{N_f} \prod_{m_\a =0}^{n_\a-1} \frac{\theta_1(\tau|(n_\a -m_\a-l-1)\frac{z}{l+1})}{\theta_1(\tau|(n_\a -m_\a)\frac{z}{l+1})}\nonumber
\\
& = (-1)^k\sum_{\vec{n} \phantom{,} \textrm{s.t}\, |\vec{n}|=k}
  \prod_{\a=1}^{N_f} \prod_{m_\a =0}^{n_\a-1} \frac{n_\a -m_\a-l-1}{n_\a -m_\a}\nonumber
\\
& =(-1)^{2k}\sum_{\vec{n} \phantom{,} \textrm{s.t}\, |\vec{n}|=k}
  \prod_{\a=1}^{N_f}{l \choose n_\a} \nonumber
\\
&= {N_f l \choose k}  \label{Witten_index_Ad}
\end{align}
where the last line follows from the fact that ${l \choose n_\a}$ is the coefficient of $x^{n_\a}$ term in a polynomial $(1+x)^l$. Thus \eqref{Witten_index_Ad} is the coefficient of $x^{n_1}x^{n_2}\cdots x^{n_{N_f}}=x^k$ term of a polynomial $(1+x)^{N_f l}$. Note that the Witten index is always positive due to the sign $(-1)^k$. This result can also be derived by turning on twisted masses for the fundamental flavors
and counting the discrete vacua of Coulomb branch.

\subsubsection{Dual theory}
The Seiberg-like dual theory is a $U(lN_f-k)$ gauge theory with matter fields of $N_f$ fundamental chiral multiplets $q$ and one adjoint chiral multiplet $Y$ and the superpotential $W=\Tr Y^{l+1}$ with the global symmetries,
\begin{equation}
\begin{array}{c|ccc}
& U(lN_f-k) & SU(N_f) & U(1)_L \\
\hline
q &  \square  &  \square& 0 \\
Y & \bold{Ad} & \bold{1} & \frac{1}{l+1}
\end{array}
\end{equation}

We would like to rewrite \eqref{EG_U(k)NfAd} in terms of the Seiberg-like dual theory. In order to obtain an expression for $U(lN_f-k)$ gauge group, we change parameters, $\tilde{n}_\a = l-n_\a$ and rearrange terms as follows.
\begin{align}
 Z(\tau,z,\xi_\a)
&= (-1)^k \sum_{\vec{n} \phantom{,} \textrm{s.t}\, |\vec{n}|=k}
  \prod_{\a,\b=1}^{N_f} \prod_{m_\a =0}^{n_\a-1} \frac{\theta_1(\tau|\xi_\a -\xi_\b +(n_\b -m_\a-l-1)\frac{z}{l+1})}{\theta_1(\tau|\xi_\a -\xi_\b +(n_\b -m_\a)\frac{z}{l+1})}
\\
&=(-1)^k \sum_{\vec{n} \phantom{,} \textrm{s.t}\, |\vec{n}|=k}
\prod_{\a,\b=1}^{N_f} \frac{\theta_1(\tau|\xi_\a -\xi_\b +(n_\b-l-1)\frac{z}{l+1})}{\theta_1(\tau|\xi_\a -\xi_\b +(n_\b )\frac{z}{l+1})}
\cdots \frac{\theta_1(\tau|\xi_\a -\xi_\b +(n_\b -n_\a-l)\frac{z}{l+1})}{\theta_1(\tau|\xi_\a -\xi_\b +(n_\b -n_\a+1)\frac{z}{l+1})} \nn
\\
&=(-1)^k \sum_{\vec{\tilde n} \phantom{,} \textrm{s.t}\, |\vec{\tilde{n}}|=lN-k}
\prod_{\a,\b=1}^{N_f} \frac{\theta_1(\tau|\xi_\a -\xi_\b +(- \tilde n_\b-1)\frac{z}{l+1})}{\theta_1(\tau|\xi_\a -\xi_\b +(-\tilde n_\b +l )\frac{z}{l+1})}
 \cdots \frac{\theta_1(\tau|\xi_\a -\xi_\b +(-\tilde{n}_\b +\tilde{n}_\a-l)\frac{z}{l+1})}{\theta_1(\tau|\xi_\a -\xi_\b +(-\tilde{n}_\b +\tilde{n}_\a+1)\frac{z}{l+1})} \nn
\\
&=(-1)^k \sum_{\vec{\tilde n} \phantom{,} \textrm{s.t}\, |\vec{\tilde{n}}|=lN-k}
\prod_{\a,\b=1}^{N_f} \prod_{\tilde{m}_\a =\tilde{n}_\a-l}^{-1}
\frac{\theta_1(\tau|\xi_\a -\xi_\b +(- \tilde n_\b + \tilde{m}_\a)\frac{z}{l+1})}{\theta_1(\tau|\xi_\a -\xi_\b +(-\tilde n_\b + \tilde{m}_\a+l+1 )\frac{z}{l+1})}\nn
\\
&=(-1)^k \sum_{\vec{\tilde n} \phantom{,} \textrm{s.t}\, |\vec{\tilde{n}}|=lN-k}
\prod_{\a,\b=1}^{N_f}
\prod_{\tilde{m}_\a =\tilde{n}_\a-l}^{\tilde{n}_\a-1} \frac{\theta_1(\tau|\xi_\a -\xi_\b +(- \tilde n_\b + \tilde{m}_\a)\frac{z}{l+1})}{\theta_1(\tau|\xi_\a -\xi_\b +(-\tilde n_\b + \tilde{m}_\a+l+1 )\frac{z}{l+1})}\nn
\\
&\qquad \qquad \qquad \qquad \qquad \qquad  \times
\prod_{\tilde{m}_\a =0}^{\tilde{n}_\a-1}\frac{\theta_1(\tau|\xi_\a -\xi_\b +(-\tilde n_\b + \tilde{m}_\a+l+1 )\frac{z}{l+1})}{\theta_1(\tau|\xi_\a -\xi_\b +(- \tilde n_\b + \tilde{m}_\a)\frac{z}{l+1})} \nn
\end{align}
The factor in the middle can be simplified as
\begin{align}
&\prod_{\a,\b=1}^{N_f} \prod_{\tilde m_\a =\tilde n_\a-l}^{\tilde n_\a-1}
\frac{\theta_1(\tau|\xi_\a -\xi_\b +(-\tilde n_\b + \tilde m_\a)\frac{z}{l+1})}{\theta_1(\tau|\xi_\a -\xi_\b +(-\tilde n_\b +\tilde m_\a+l+1)\frac{z}{l+1})} \label{EG_AdRelation1}
\\
&=\prod_{\a,\b=1}^{N_f} \frac{\theta_1(\tau|\xi_\a -\xi_\b +(-\tilde n_\b + \tilde n_\a - l)\frac{z}{l+1})}{\theta_1(\tau|\xi_\a -\xi_\b +( - \tilde n_\b + \tilde n_\a+1)\frac{z}{l+1})}\times\cdots\times
\frac{\theta_1(\tau|\xi_\a -\xi_\b +(-\tilde n_\b +\tilde n_\a-1)\frac{z}{l+1})}{\theta_1(\tau|\xi_\a -\xi_\b +(-\tilde n_\b +\tilde n_\a+l)\frac{z}{l+1})}\nn
\\
&=(-1)^{lN_f^2}\prod_{\a,\b=1}^{N_f} \frac{\theta_1(\tau|-\xi_\a +\xi_\b +(\tilde n_\b - \tilde n_\a+l)\frac{z}{l+1})}{\theta_1(\tau|\xi_\a -\xi_\b +(-\tilde n_\b +\tilde n_\a+1)\frac{z}{l+1})}\times\cdots\times
\frac{\theta_1(\tau|-\xi_\a +\xi_\b +(\tilde n_\b -\tilde n_\a+1)\frac{z}{l+1})}{\theta_1(\tau|\xi_\a -\xi_\b +(-\tilde n_\b + \tilde n_\a+l)\frac{z}{l+1})}\nn
\\
&=(-1)^{lN_f^2}\prod_{\a,\b=1}^{N_f} \frac{\theta_1(\tau|-\xi_\b +\xi_\a +(\tilde n_\a - \tilde n_\b+l)\frac{z}{l+1})}{\theta_1(\tau|\xi_\a -\xi_\b +(-\tilde n_\b +\tilde n_\a+1)\frac{z}{l+1})}\times\cdots\times
\frac{\theta_1(\tau|-\xi_\b +\xi_\a +(\tilde n_\a -\tilde n_\b+1)\frac{z}{l+1})}{\theta_1(\tau|\xi_\a -\xi_\b +(-\tilde n_\b + \tilde n_\a+l)\frac{z}{l+1})}\nn
\\
&=(-1)^{lN_f^2}\nn
\end{align}
where we reverse all the signs of  theta functions in the numerator using $\theta_1(\tau|-a)=-\theta_1(\tau|a)$ at the second line and exchange $\a$ and $\b$ dummy indices in the numerator at the third line, then all numerators and denominators are canceled against each other.
Then the elliptic genus becomes
\begin{align}
 Z(\tau,z,\xi_\a)
&=(-1)^{lN_f-k}\sum_{\vec{\tilde n} \phantom{,} \textrm{s.t}\, |\vec{\tilde{n}}|=lN_f -k}\prod_{\a, \b=1}^{N_f}\prod_{\tilde m_\a =0}^{\tilde{n}_\a-1} \frac{\theta_1(\tau| -\xi_\a +\xi_\b +(\tilde{n}_\b-\tilde m_\a-l-1)\frac{z}{l+1})}{\theta_1(\tau|-\xi_\a +\xi_\b +(\tilde{n}_\b-\tilde m_\a)\frac{z}{l+1})} \label{EG_U(k)NfAd_sub2}
\end{align}
where we reversed all the signs of  theta functions using $\theta_1(\tau|-a)=-\theta_1(\tau|a)$. This is nothing but the elliptic genus of the $U(lN_f-k)$ gauge theory

\subsection{$U(k)$ with $N_f$ fundamentals, $N_a$ anti-fundamentals and one adjoint}
We consider a $U(k)$ gauge theory with $N_f$ chiral multiplets in fundamental representation, $N_a$ chiral multiplets in anti-fundamental representation and one chiral multiplet in adjoint representation and the superpotential of the form, $W=\Tr X^{l+1}$.
The charges of flavor symmetries and a left-moving R-symmetry $U(1)_L$ are
\begin{equation}
\begin{array}{c|ccccc}
& U(k) & SU(N_f) & SU(N_a) & U(1)_a & U(1)_L \\
\hline
Q & \square & \overline\square & \bold{1} & 1 & 0 \\
\tilde Q & \overline\square & \bold{1} & \square & 1 & 0\\
X & \bold{Ad} & \bold{1} & \bold{1} & 0 & \frac{1}{l+1}
\end{array}
\end{equation}

Seiberg-like dual theory is a $U(lN_f-k)$ gauge theory with matter fields of $N_f$ fundamentals $q_\a$, $N_a$ anti-fundamentals $\tilde q^\g$ and one adjoint $Y$, and $lN_fN_a$ singlets $M_j^{\a \g}$, $j=0,...,l-1$, $\a=1,...,N_f$, $\g=1,...,N_a$. It has the superpotential, $W=\Tr Y^{l+1}+ M_j\tilde q Y^{l-1-j} q$ which fixes,  together with the identification $M_j\leftrightarrow Q X^j\wtilde Q$, charges
\begin{equation}
\begin{array}{c|ccccc}
& U(lN_f-k) & SU(N_f) & SU(N_a) & U(1)_a & U(1)_L \\
\hline
q & \square & \square & \bold{1} & -1 & \frac{1}{l+1} \\
\tilde q & \overline\square & \bold{1} & \overline \square & -1 & \frac{1}{l+1}\\
M_j & \bold{1} & \overline \square & \square & 2 & \frac{j}{l+1} \\
Y & \bold{Ad} & \bold{1} & \bold{1} & 0 & \frac{1}{l+1}
\end{array}
\end{equation}

One-loop determinant of the $U(k)$ gauge theory is
\begin{multline}\label{EG_1loop_NfNaAd}
Z_\text{1-loop} = (-1)^k \frac1{k!} \bigg( \frac{2\pi \eta(q)^3}{\theta_1(q,y^{-1})} \bigg)^k \bigg( \prod_{i\neq j}^k \frac{\theta_1(\tau|u_i - u_j)}{\theta_1(\tau|u_i - u_j - z)} \bigg)  \bigg( \prod_{i,j=1}^k \frac{\theta_1(\tau|u_i - u_j + \frac{1}{l+1}z - z)}{\theta_1(\tau|u_i - u_j + \frac{1}{l+1}z)} \bigg)
 \\
\times
 \bigg(\prod_{i=1}^k \prod_{\alpha=1}^{N_f} \frac{\theta_1(\tau| u_i - \xi_\alpha +\chi - z)}{\theta_1(\tau|u_i - \xi_\alpha + \chi)}\bigg)
 \bigg(\prod_{i=1}^k \prod_{\gamma=1}^{N_a} \frac{\theta_1(\tau| - u_i +  \eta_\gamma + \chi - z)}{\theta_1(\tau|- u_i + \eta_\gamma + \chi)}\bigg) \, \dd^ku \;.
\end{multline}
where $\xi_\a$, $\eta_\g$, $\chi$ are holonomies for the $SU(N_f) \times SU(N_a)\times U(1)_a$ flavor symmetry. Single-valuedness of the one-loop determinant requires $y^{N_f-N_a}=1$.

Let us assume $N_f\geq N_a$. The JK residue comes from the same pole configurations as the case without anti-fundamentals because the JK residue can be evaluated by poles associated with positive charges so that the poles from anti-fundamentals do not contribute.
\begin{align}\hspace{-1cm}\label{EG_U(k)N_fN_aAd}
 Z  & = (-1)^k \sum_{\vec{n} \phantom{,} \textrm{s.t}\, |\vec{n}|=k}  \prod_{\a,\b=1}^{N_f} \prod_{m_\a =0}^{n_\a-1}
\frac{\theta_1(\tau|\xi_\a -\xi_\b +(n_\b -m_\a-l-1)\frac{z}{l+1})}{\theta_1(\tau|\xi_\a -\xi_\b +(n_\b -m_\a)\frac{z}{l+1})}
\times\\
&\qquad\qquad\qquad\qquad
\prod_{\a=1}^{N_f}\prod_{\g=1}^{N_a} \prod_{m_\a =0}^{n_\a-1}
\frac{\theta_1(\tau|-\xi_\a +\eta_\g +2\chi+(m_\a-l-1)\frac{ z}{l+1}))}{\theta_1(\tau|-\xi_\a +\eta_\g +2\chi+m_\a\frac{ z}{l+1})}  \nn ~.
\end{align}

The first line of \eqref{EG_U(k)N_fN_aAd} is the same as that of the theory without anti-fundamentals whose contributions appear only in the second line. It can be written as
\begin{align}
&\prod_{\a=1}^{N_f}\prod_{\g=1}^{N_a}\prod_{m_\a=0}^{n_\a-1}\frac{\theta_1(\tau|-\xi_\a +\eta_\g +2\chi+\frac{ (m_\a-l-1)z}{l+1})}{\theta_1(\tau|-\xi_\a +\eta_\g+2\chi +\frac{m_\a z}{l+1})} \label{EG_U(k)NfNaAd_1}
\\
&=\left(\prod_{\a=1}^{N_f}\prod_{\g=1}^{N_a}\prod_{m_\a=0}^{l-1}\frac{\theta_1(\tau| -\xi_\a+\eta_\g +2\chi +\frac{(m_\a-l-1)z}{l+1})}{\theta_1(\tau|-\xi_\a+\eta_\g+2\chi+\frac{m_\a z}{l+1})}\right)
\left(\prod_{\a=1}^{N_f}\prod_{\g=1}^{N_a} \prod^{l-1}_{m_\a=n_\a}\frac{\theta_1(\tau|-\xi_\a+\eta_\g+2\chi+\frac{m_\a z}{l+1})}{\theta_1(\tau| -\xi_\a+\eta_\g +2\chi +\frac{(m_\a-l-1)z}{l+1})} \right)
\nn
\\
&=\left(\prod_{\a=1}^{N_f}\prod_{\g=1}^{N_a}\prod_{j=0}^{l-1}\frac{\theta_1(\tau| -\xi_\a+\eta_\g +2\chi +\frac{(j-l-1)z}{l+1})}{\theta_1(\tau|-\xi_\a+\eta_\g+2\chi+\frac{jz}{l+1})}\right)
\left(\prod_{\a=1}^{N_f}\prod_{\g=1}^{N_a} \prod_{\tilde m_\a=0}^{\tilde{n}_\a-1}\frac{\theta_1(\tau|\xi_\a-\eta_\g-2\chi+\frac{(\tilde m_\a-l+1)z}{l+1})}{\theta_1(\tau| \xi_\a-\eta_\g -2\chi +\frac{(\tilde m_\a+2)z}{l+1})} \right)
\nn
\end{align}
where $\tilde n_\a=-n_\a+l$, $\tilde m_\a = -m_a+l-1$ and we used $\theta_1(\tau|-a)=-\theta_1(\tau|a)$. Note that the second line is possible because $0\leq n_\a \leq l$. When $n_\a=l$, $\tilde n_\a =0$ the second factors in the second and third lines of \eqref{EG_U(k)NfNaAd_1} are 1. Therefore, the elliptic genus can be written as
\begin{align}\hspace{-1.5cm}
Z&= (-1)^k\sum_{\vec{n} \phantom{,} \textrm{s.t}\, |\vec{n}|=k}  \prod_{\a,\b=1}^{N_f} \prod_{m_\a =0}^{n_\a-1}
\frac{\theta_1(\tau|\xi_\a -\xi_\b +\frac{(n_\b -m_\a-l-1)z}{l+1})}{\theta_1(\tau|\xi_\a -\xi_\b +\frac{(n_\b -m_\a)z}{l+1})}
\prod_{\a=1}^{N_f}\prod_{\g=1}^{N_a} \prod_{m_\a =0}^{n_\a-1}
\frac{\theta_1(\tau|-\xi_\a +\eta_\g +2\chi+\frac{(m_\a-l-1) z}{l+1})}{\theta_1(\tau|-\xi_\a +\eta_\g +2\chi+\frac{m_\a z}{l+1})}
\nn
\\
&=(-1)^{lN_f-k}\sum_{\vec{\tilde n} \phantom{,} s.t |\vec{\tilde{n}}|=lN_f-k}  \prod_{\a, \b=1}^{N_f}\prod_{\tilde m_\a =0}^{\tilde n_\a-1}
   \frac{\theta_1(\tau| -\xi_\a +\xi_\b +\frac{(\tilde n_\b-\tilde m_\a-l-1)z}{l+1})}{\theta_1(\tau|-\xi_\a +\xi_\b +\frac{(\tilde n_\b-\tilde m_\a)z}{l+1})}
\label{EG_U(k)NfNaAd_dual}
\\
   &\qquad \times
\prod_{\a=1}^{N_f}\prod_{\g=1}^{N_a} \prod_{\tilde m_\a =0}^{\tilde n_\a-1}
\frac{\theta_1(\tau|\xi_\a-\eta_\g - 2\chi+\frac{(\tilde m_\a-l+1)z}{l+1})}{\theta_1(\tau|\xi_\a -\eta_\g-2\chi +\frac{(\tilde m_\a+2)z}{l+1})}
\prod_{j=0}^{l-1}\frac{\theta_1(\tau| -\xi_\a+\eta_\g + 2\chi+\frac{(j-l-1)z}{l+1})}{\theta_1(\tau|-\xi_\a+\eta_\g+2\chi+\frac{jz}{l+1})}\nn
\end{align}
The second expression is the elliptic genus of the $U(lN_f-k)$ gauge theory. The first fraction in the third line comes from $q_\a \tilde q_\g Y^{m_\a}$, $\mathcal Q (\wbar{ q_\a \tilde q_\g Y^{m_\a}})$ and their complex conjugates. The second fraction in the third line comes from $M_j$, $\mathcal Q (\wbar{M}_j)$ and their complex conjugates. All contributions coming from $q_\a \tilde q_\g Y^{m_\a}$ are canceled out. This can be seen from \eqref{EG_U(k)NfNaAd_1}. The first and the second fractions in the second line of \eqref{EG_U(k)NfNaAd_1} are identified as contributions from $M_j$ and $q_\a \tilde q_\g Y^{m_\a}$ respectively. All the second fractions are canceled out with factors $m_\a=n_\a,\ldots,l-1$ of the first fraction. When $k<l$ we have $0\leq n_\a \leq k$ so only $M_i$, $i=0,\ldots, k-1$ contributions $\left(\prod_{\a=1}^{N_f}\prod_{\g=1}^{N_a}\prod_{j=0}^{k-1}\frac{\theta_1(\tau| -\xi_\a+\eta_\g +2\chi +\frac{(j-l-1)z}{l+1})}{\theta_1(\tau|-\xi_\a+\eta_\g+2\chi+\frac{jz}{l+1})}\right)$ remain. When $k\geq l$ the non-trivial contributions are $0\leq n_\a \leq l$ so all $M_j$, $j=0,\ldots, l-1$ contribute.

\subsubsection{Characteristic of the theory}

Interestingly, the elliptic genus has non-trivial $q$ dependence even though the single-valuedness condition $y^{N_f-N_a}=1$ is imposed, which is different from $N_a=0$ case. The nontrivial $q$ dependence comes from $M_j=QX^j \tilde{Q}$, $j=0,\ldots, \textrm{min}(l-1,k-1)$. Due to $M_j$, the  $q^0$ terms representing ground state contributions also depend on the flavor symmetry fugacities, while  the elliptic genus of the theories without anti-fundamental matter fields are independent of the flavor fugacities.
This is related to the fact that the theory has additional Higgs branch, parametrized by $M_j$ as well as discrete vacua.
In fact, the special case of $U(1)$ theory with $(N_f, N_a)$ was analyzed at \cite{Hanany:1997vm}. They considered $U(1)$ gauge theory with $N_f$ fundamental chiral fields $Q^\alpha$ of charge 1 and $N_a$ anti-fundamental chiral fields $\tilde Q^\beta$ of charge $-1$ and all chiral fields have twisted masses. When one pair of $Q^\alpha$ and $\tilde Q^\beta$ have the same twisted mass and the other chiral fields have generic twist masses the theory has $\max(N_f-1, N_a-1)$ massive, discrete Coulomb vacua and the vacua of a sigma model on one-dimensional non-compact complex manifold parametrized by $Q^\alpha$ and $\tilde Q^\beta$ for some $\alpha$ and $\beta$. When we turn off the twisted masses, the theory has $|N_f-N_a|$ discrete vacua and $N_f N_a$ dimensional Higgs
branch. This is consistent with the elliptic genus result. As an example,  consider $U(1)$ theory with $N_f$ chiral fields of charge 1 and one chiral field of charge $-1$. The elliptic genus is given by
\begin{align}
Z=\sum_{\alpha=1}^{N_f}\left(\prod_{\beta\neq \alpha} \frac{\theta_1(\tau|\xi_\alpha -\xi_\beta -z)}{\theta_1(\tau|\xi_\alpha -\xi_\beta)}\right) \frac{\theta_1(\tau|-\xi_\alpha +2\chi -z)}{\theta_1(\tau|-\xi_\alpha +2\chi)}
\end{align}
The ground states contribution is obtained as
\begin{align}
Z(q\rightarrow 0) = y^{-(N_f-2)/2}(1+y+\cdots+y^{N_f-2}) + \prod_{\alpha=1}^{N_f} y^{-1/2}d^{2}\frac{1-ya_\alpha d^{-2}}{1-a_\alpha^{-1} d^2}
\end{align}
where the first term is the contribution of $(N_f-1)$ massive vacua and the second term comes from  $N_f$ dimensional Higgs vacua.
Existence of both discrete vacua and continuous Higgs branch persists for generic $U(n)$ theory with $N_f \neq N_a$ with one adjoint.
This can be confirmed by the effective action analysis similar to \cite{Hanany:1997vm} and the elliptic genus result is consistent with it.

The mesonic fields $(M_j)^{\alpha\beta}$ parameterizing the Higgs branch are not free fields in general. If $k < N_a$, this can be easily
 seen since maximum rank of $(M_j)^{\alpha\beta}$ is constrained by the gauge group rank. For example, $M_0$ fields can be represented as a $N_f\times N_a$ matrix. If $k<N_a$ the rank of $M_0$ matrix is $k$ so that we have the usual rank condition,
\begin{align}
\epsilon_{\alpha_1\ldots \alpha_{k+1}}\epsilon_{\beta_1\ldots \beta_{k+1}}M_0^{\alpha_1\beta_1}\cdots M_0^{\alpha_{k+1}\beta_{k+1}}=0
\end{align}
where $\alpha_i\in I$, $\beta_i\in J$ and $I$, $J$ are $(k+1)$-combinations of $\{1,\ldots,N_a\}$.

The adjoint field is constrained by the characteristic equation of $k\times k$ matrix.
\begin{align}
X^k - (\Tr X) X^{k-1} +\cdots + (-1)^k (\det X) I_k =0
\end{align}
where $I_k$ is $k\times k$ identity matrix. By contracting the equation with $Q$ and $\tilde Q$, we have
\begin{align}
M_k - (\Tr X) M_{k-1} +\cdots + (-1)^k (\det X) M_0 =0~.
\end{align}
This shows that if $k<l$ only $M_0$, $\ldots$, $M_{k-1}$ are independent. If $k\geq l$ then $M_k=M_{k-1}=\cdots=M_{l}=0$ and remaining fields are constrained by the equation.
Also computation of the elliptic genus shows that the mesonic fields are not free fields generically.
\footnote{As an example, $U(1)$ with $N_f=2, N_a=1$ the elliptic genus is given by
\begin{align}\hspace{-1cm}
Z= 1+ \frac{d^{4}}{y}\frac{(1-ya_1d^{-2})(1-ya_2d^{-2})}{(1-a_1^{-1}d^2)(1-a_2^{-1}d^2)}+ q\left( \frac{d^2(1-y)(a_1+a_2)(1-ya_1d^{-2})(1-ya_2d^{-2})(1-d^2(1-y)(a_1+a_2)-yd^4)}{y^2(1-a_1^{-1}d^2)(1-a_2^{-1}d^2)} \right) +O(q^2).
\end{align}
where $O(q)$ term clearly shows that it is not just contribution of two free chiral fields.}

The exception occurs when $lN_f=k$ with $N_f \geq N_a>0$. The corresponding theory is dual to the theory of $N_f N_a$ free chiral fields.

\subsection{$U(k)$ with $N$ pairs of fundamentals/anti-fundamentals and one adjoint}
Here we consider the duality of the superconformal field theories.
We consider the A theory,  $U(k)$ gauge theory with $N$ chiral multiplets in fundamental representation, $N$ chiral multiplets in anti-fundamental representation and one chiral multiplet in adjoint representation and the superpotential of the form, $W=\Tr X^{l+1}$.
The charges of the flavor symmetries and a left-moving R-symmetry $U(1)_L$ are
\begin{equation}
\begin{array}{c|ccccc}
& U(k) & SU(N) & SU(N) & U(1)_a & U(1)_L \\
\hline
Q & \square & \overline\square & \bold{1} & 1 & 0 \\
\tilde Q & \overline\square & \bold{1} & \square & 1 & 0\\
X & \bold{Ad} & \bold{1} & \bold{1} & 0 & \frac{1}{l+1}
\end{array}
\end{equation}
The dual theory, which is called B theory is
 $U(lN-k)$ gauge theory with matter fields of $N$ pairs of fundamental, anti-fundamental and one adjoint, and $lN^2$ singlets $M_j^{\a\b}$, $j=0,...,l-1$.
It has the superpotential, $W=\Tr Y^{l+1}+ \sum_{j=0}^{l-1}M_j\tilde q Y^{l-1-j} q$ which fixes,  together with the identification $M_j\leftrightarrow Q X^j\wtilde Q$, charges
\begin{equation}
\begin{array}{c|ccccc}
& U(lN-k) & SU(N) & SU(N) & U(1)_a & U(1)_L \\
\hline
q & \square & \square & \bold{1} & -1 & \frac{1}{l+1} \\
\tilde q & \overline\square & \bold{1} & \overline \square & -1 & \frac{1}{l+1}\\
M_j & \bold{1} & \overline \square & \square & 2 & \frac{j}{l+1} \\
Y & \bold{Ad} & \bold{1} & \bold{1} & 0 & \frac{1}{l+1}
\end{array}
\end{equation}

We will see chiral ring generators of the theory by analyzing elliptic genus of the theories. Chiral ring generators coming from the adjoint field are $\Tr X^i\,/\,\Tr Y^i$, $i=1,\ldots, \min(l-1,k, lN-k)$, which are constrained by $F$-term condition and the characteristic equations of the adjoint fields. The A and B theories have different gauge groups so the characteristic equations are different. To have a consistent chiral ring with the duality, a characteristic equation of one theory is a quantum constraint of the dual theory if the constraint is stronger than the other. Chiral ring generators coming from the mesonic operators are $QX^j\tilde Q \,/\, M_j$, $j=0,\ldots, \min(l-1,k-1)$. Unlike the adjoint operators, the mesonic operators are constrained only by the A theory classical equations.

The elliptic genera of A theory and B theory are given by
\begin{align}
Z^A&= (-1)^k
\sum_{\vec{n} \phantom{,} \textrm{s.t}\, |\vec{n}|=k}  \prod_{\a,\b=1}^{N} \prod_{m_\a =0}^{n_\a-1}
\frac{\theta_1(\tau|\xi_\a -\xi_\b +\frac{(n_\b -m_\a-l-1)z}{l+1})}{\theta_1(\tau|\xi_\a -\xi_\b +\frac{(n_\b -m_\a)z}{l+1})}
\frac{\theta_1(\tau|-\xi_\a +\eta_\b +2\chi+\frac{(m_\a-l-1) z}{l+1})}{\theta_1(\tau|-\xi_\a +\eta_\b +2\chi+\frac{m_\a z}{l+1})}\nonumber
\\
&=(-1)^{lN-k}\sum_{\vec{\tilde n} \phantom{,} s.t |\vec{\tilde{n}}|=lN-k}  \prod_{\a, \b=1}^{N}\prod_{\tilde m_\a =0}^{\tilde n_\a-1}
   \frac{\theta_1(\tau| -\xi_\a +\xi_\b +\frac{(\tilde n_\b-\tilde m_\a-l-1)z}{l+1})}{\theta_1(\tau|-\xi_\a +\xi_\b +\frac{(\tilde n_\b-\tilde m_\a)z}{l+1})}
\frac{\theta_1(\tau|\xi_\a-\eta_\b-2\chi+\frac{(\tilde m_\a-l+1)z}{l+1})}{\theta_1(\tau|\xi_\a -\eta_\b-2\chi +\frac{(\tilde m_\a+2)z}{l+1})}
\nn
\\
   &\qquad \times
  \prod_{j=0}^{l-1}\frac{\theta_1(\tau| -\xi_\a+\eta_\b +2\chi+\frac{(j-l-1)z}{l+1})}{\theta_1(\tau|-\xi_\a+\eta_\b+2\chi+\frac{jz}{l+1})}
\end{align}
where $\xi_\a$, $\eta_\b$, $\chi$ are holonomies for $SU(N) \times SU(N)\times U(1)_a$ flavor symmetry.
The single-valuedness condition of the one-loop determinant does not require any condition on $y$ and this is consistent with the fact that the theories have no R-symmetry anomaly.

\subsubsection{Modular Property}
Let us check the central charge of the theories. The central charge, $c$, can be obtained from the modular property of the elliptic genus.
\begin{align}
Z\left(-\frac{1}{\t}, \frac{z}{\t}, \frac{u_i}{\t}\right)=e^{\frac{\pi i}{\t}\left(-2A^iu_iz+\frac{c}{3}z^2\right)}Z(\t,z,u_i)
\end{align}
where $A^i$ is the t'~Hooft anomaly between the left-moving R-symmetry and flavor symmetry $K^i$. It can be computed using the modular property of the theta function,
\begin{align}
\theta_1 \Big( - \frac1\tau \Big| \frac z\tau \Big) = -i \, \sqrt{-i \tau} \, e^{\pi i z^2/\tau} \, \theta_1(\tau| z)
\end{align}
The modular transformation of the elliptic genus of the A-theory is
\begin{align}\hspace{-1cm}
Z^A\left(-\frac{1}{\t}, \frac{z}{\t}, \frac{\xi_\a}{\t},\frac{\eta_\b}{\t},\frac{\chi}{\t}\right)
&=\sum_{\vec{n} \phantom{,} \textrm{s.t}\, |\vec{n}|=k}  \prod_{\a,\b=1}^{N} \prod_{m_\a =0}^{n_\a-1}
e^{\frac{\pi i}{\t}\left(-2z(\xi_\a-\xi_\b) +z^2\left(1-\frac{2(n_\b-m_\a)}{l+1}\right)\right)}
e^{\frac{\pi i}{\t}\left(-2z(-\xi_\a+\eta_\b+2\chi) +z^2\left(1-\frac{2m_\a}{l+1}\right)\right)} \nn
\\
&\qquad \qquad \qquad
\frac{\theta_1(\tau|\xi_\a -\xi_\b +\frac{(n_\b -m_\a-l-1)z}{l+1})}{\theta_1(\tau|\xi_\a -\xi_\b +\frac{(n_\b -m_\a)z}{l+1})}
\frac{\theta_1(\tau|-\xi_\a +\eta_\b +2\chi+\frac{(m_\a-l-1) z}{l+1})}{\theta_1(\tau|-\xi_\a +\eta_\b +2\chi+\frac{m_\a z}{l+1})}
\end{align}
It can be written as
\begin{align}
Z^A\left(-\frac{1}{\t}, \frac{z}{\t}, \frac{\xi_\a}{\t},\frac{\eta_\b}{\t},\frac{\chi}{\t}\right)
&=e^{\frac{\pi i}{\t}\left(-4zkN\chi+\frac{c}{3}z^2\right)}Z^A(\t,z,\xi_\a,\eta_\b,\chi)
\end{align}
where $\sum_{\b=1}^N\eta_\b=\sum_{\b=1}^N\xi_\b=0$ has been used because they are $SU(N)$ flavor holonomies and the central charge is
\begin{align}
\frac{c}{3}=\sum_{\a,\b=1}^N\sum_{m_\a=0}^{n_\a-1}\left(2-\frac{2n_\b}{l+1}\right)=2k\left(N-\frac{k}{l+1}\right)=2kN-k^2+k^2\left(1-\frac{2}{l+1}\right)
\end{align}
where $\sum_{\b=1}^Nn_\b=k$.
It is equal to the central charge of the GLSM, i.e. $2kN$ from chiral multiplets of $(Q,\tilde Q)$, $-k^2$ from $U(k)$ vector multiplet and $k^2\left(1-\frac{2}{l+1}\right)$ from adjoint chiral multiplet. The central charge of the B-theory computed from the elliptic genus is
\begin{align}
\frac{c}{3}=2(lN-k)N\left(1-\frac{2}{l+1}\right)-(lN-k)^2+(lN-k)^2\left(1-\frac{2}{l+1}\right)+N^2\sum_{j=0}^{l-1}\left(1-\frac{2j}{l+1}\right)
\end{align}
which is equal to the central charge of the A-theory.

 Alternatively when a two dimensional theory has at least $\mathcal N=(0,2)$ supersymmetry the central charge can be obtained from the 't Hooft anomaly of the R-symmetry. Non-conformal R-symmetry can be mixed with flavor symmetries while the superconformal R-symmetry is identified by requiring that it does not have a cross anomaly with any global symmetry. Furthermore, the superconformal right(left)-moving R-current should not be mixed with non-(anti-)holomorphic currents of flavor symmetries \cite{Benini:2013cda} where (anti-)holomorphic means right(left)-mover. The flavor symmetries of the theories of interest are $SU(N)\times SU(N) \times U(1)_a$ which are not purely left-moving or purely right-moving. In this situation the central charge of $\mathcal N=(2,2)$ superconformal theories can be computed as
\begin{align}
\frac{c}{3}
=\sum_{i\text{: Weyl fermions}}\gamma^3 R_iR_i = \sum_{i}\left(\frac{R_i}{2}-1\right)^2-\left(\frac{R_i}{2}\right)^2
=\sum_{j\text{: Dirac fermions}}(1-R_j)
\end{align}
where $ \gamma^3 $ is the chirality matrix taking 1 on right-movers and $-1$ on left-movers, $R_i$ is the right-moving R-charge of a Weyl fermion $ \psi_\pm $, $R_j-1$ is the vector R-charge of a Dirac fermion $ \binom{\psi_+}{\psi_-} $. The vector R-charge is the linear combination of
left-moving R-charge and right-moving R-charge, which is the same as the usual R-charge of $\mathcal N=(2,2)$ multiplets. For example, a theory of a free chiral multiplet whose fermionic component field has vector R-charge $-1$ has the central charge 1. Let us compute the central charge of the theory. The fundamental and anti-fundamental chiral multiplets have vector R-charge 0 so fermion fields have vector R-charge -1. The gaugino fields have the vector R-charge 1. The adjoint chiral multiplet have vector R-charge $ \frac{2}{n+1} $ so its fermion field has vector R-charge $ \frac{2}{n+1} -1$. Therefore, the central charge is given by
\begin{align}
\frac{c}{3}=2kN-k^2+k^2\left(1-\frac{2}{l+1}\right)~,
\end{align}
which is the same as the one obtained from the modular property of the elliptic genus.

\subsubsection{Chiral Primaries}
In this subsection, we check the matching of the chiral primaries between the dual theory.
By taking a limit $\tau \rightarrow i\infty$ the elliptic genus becomes
\begin{align}
\lim_{\tau \rightarrow i\infty} Z^A(\t,z,\xi_\a,\eta_\b,\chi)
=(-1)^k\sum_{\vec{n} \phantom{,} \textrm{s.t}\, |\vec{n}|=k}  \prod_{\a,\b=1}^{N} \prod_{m_\a =0}^{n_\a-1}
& (-1)y^{\frac{n_\b-m_\a}{l+1}-\frac{1}{2}}(a_\a a_\b^{-1})\frac{1-y^{\frac{l+1-n_\b+m_\a}{l+1}}a_\a^{-1} a_\b }{1-y^{\frac{n_\b-m_\a}{l+1}}a_\a a_\b^{-1}}\nn
\\
&\times (-1)y^{\frac{m_\a}{l+1}-\frac{1}{2}}d^2(a_\a^{-1} b_\b)\frac{1-y^{\frac{l+1-m_\a}{l+1}}d^{-2}a_\a b_\b^{-1}}{1-y^{\frac{m_\a}{l+1}}d^2 a_\a^{-1} b_\b}
\end{align}
where $y=e^{2i\pi z}$, $a_\a=e^{2i\pi \xi_\a}$, $b_\a=e^{2i\pi \eta_\a}$, $d=e^{2i\pi \chi}$.
It is simplified as
\begin{align} \label{Ground_U(k)NNAd}
Z^A(i\infty,z,\xi_\a,\eta_\b,\chi)
&=(-1)^ky^{\frac{k^2}{l+1}-kN}d^{2kN}
\sum_{\vec{n} \phantom{,} \textrm{s.t}\, |\vec{n}|=k}  \prod_{\a,\b=1}^{N} \prod_{m_\a =0}^{n_\a-1}
\frac{1-y^{\frac{l+1-n_\b+m_\a}{l+1}}a_\a^{-1} a_\b}{1-y^{\frac{n_\b-m_\a}{l+1}}a_\a a_\b^{-1}}
\frac{1-y^{\frac{l+1-m_\a}{l+1}}d^{-2}a_\a b_\b^{-1}}{1-y^{\frac{m_\a}{l+1}}d^2a_\a^{-1} b_\b}
\end{align}
where we used $\prod_{\a=1}^{N}a_\a=\prod_{\a=1}^{N}b_\a=1$.

The spectral flow relation $Z(q,y)=-a y^{-\frac{c}{6}}\CI(q,q^{-\frac{1}{2}}y)$ and the central charge obtained in the previous subsection can be used to identify the contributions of chiral primaries from the ground states contribution \eqref{Ground_U(k)NNAd}. Up to the factor $(-1)^kd^{2kN}$ the chiral primary contribution is given by
\begin{align}\label{CR_U(k)NNAd}
\CI^A_{CR}(q,y,a_\a,b_\b,d)=\sum_{\vec{n} \phantom{,} \textrm{s.t}\, |\vec{n}|=k}  \prod_{\a,\b=1}^{N} \prod_{m_\a =0}^{n_\a-1}
\frac{1-(yq^{1/2})^{\frac{l+1-n_\b+m_\a}{l+1}}a_\a^{-1} a_\b}{1-(yq^{1/2})^{\frac{n_\b-m_\a}{l+1}}a_\a a_\b^{-1}}
\frac{1-(yq^{1/2})^{\frac{l+1-m_\a}{l+1}}d^{-2}a_\a b_\b^{-1}}{1-(yq^{1/2})^{\frac{m_\a}{l+1}}d^2a_\a^{-1} b_\b}
\end{align}
where $Z(q\rightarrow 0,y,a_\a,b_\b,d)=(-1)^ky^{-c/6}d^{2kN}\CI_{CR}(q,q^{-1/2}y,a_\a,b_\b,d)$. \eqref{CR_U(k)NNAd} also can be obtained directly from the superconformal index with NS-NS boundary conditions \cite{Gadde:2013ftv}.
The factors $1/(1-(yq^{1/2})^{\frac{m_\a}{l+1}}d^2a_\a^{-1} b_\b)$ are contributions of operators of the form $Q_\a X^{m_\a}\tilde Q_\b$ and $1-(yq^{1/2})^{\frac{l+1-m_\a}{l+1}}d^{-2}a_\a b_\b^{-1}$ are contributions of $\mathcal Q_-(\wbar{Q_\a X^{m_\a}\tilde Q_\b})$ where $\mathcal Q_-$ is the left-moving supercharge.

The denominator of the chiral primary contribution corresponds to bosonic generators of the chiral ring.
We have numerically checked that \eqref{CR_U(k)NNAd} takes the form of
\begin{align}
\CI_{CR}(x,a_\a,b_\b,d)= \left(\prod_{\a,\b=1}^{N}\prod_{j=0}^{\min(l, k)-1}\frac{1}{1-x^{\frac{j}{l+1}}d^2a_\a^{-1} b_\b}\right) N(x,a_\a,b_\b,d)
\end{align}
where $x=yq^{1/2}$ and $N(x,a_\a,b_\b,d)$ is a polynomial in $x$, which starts with 1. The chiral ring generators are adjoint operators, $\Tr X^{j}$, $j=1,\ldots,\min(l-1,k,lN-k)$ and mesonic operators $Q_\a X^{j}\tilde Q_\b$ where $j=0,\ldots, \min(l-1,k-1)$. The constraint on the power of $X$ comes from $F$-flatness condition, $X^l=0$ and the characteristic equation of the adjoint field, which makes $Q_\a X^{j}\tilde Q_\b$, $j\geq k$ not linearly independent. The adjoint contribution is finite because of the superpotential so it appears in the numerator.
$N(x,a_\a,b_\b,d)$ consists of terms corresponding to identity, adjoint contribution, fermion contributions $\mathcal Q_-(M_j)$ and relations of chiral primaries. Contributions of $\mathcal Q_-(M_j)$ can be distinguished from that of relations because global charges are different. The relations come from the rank of mesonic matrices $(M_j)_{\a\b}=Q_\a X^j\tilde Q_\b$ and $F$-term condition, $X^l=0$. One example is $\det M=0$ for $k<N$, which is encoded as $-\prod_{\a,\b=1}^{N}a_\a^{-1} b_\b$ in the numerator. For example, when $N=2$, $k=1$ the operators $(M_0)_{11}(M_0)_{22}$ and $(M_0)_{12}(M_0)_{21}$ have the same flavor symmetry fugacity, $a_1^{-1} a_2^{-1} b_1  b_2$ and are linearly dependent. Thus one linear combination of the operators is canceled by the term $- a_1^{-1} a_2^{-1} b_1  b_2$ in the numerator.
Another example is $(M_0)_{\a\b}(\Tr X)^2  - (M_1)_{\a\b}\Tr X=0$ for $U(2)$ gauge theory with $X^2=0$, which corresponds to $-x^{\frac{2}{3}}a_\a^{-1} b_\b$.

In the limit $\tau\rightarrow i\infty$, the elliptic genus of the dual theory becomes
\begin{align}
&\lim_{\tau \rightarrow i\infty} Z^B(\t,z,\xi_\a,\eta_\b,\chi)\nn
\\
&=(-1)^{lN-k}\sum_{\vec{\tilde n} \phantom{,} \textrm{s.t}\, |\vec{\tilde n}|=lN-k}
 \prod_{\a,\b=1}^{N}\left( \prod_{\tilde m_\a =0}^{\tilde n_\a-1}
(-1) y^{\frac{\tilde n_\b-\tilde m_\a}{l+1}-\frac{1}{2}}(a_\a^{-1} a_\b)\frac{1-y^{\frac{l+1-\tilde n_\b + \tilde m_\a}{l+1}}a_\a a_\b^{-1}}{1-y^{\frac{\tilde n_\b-\tilde m_\a}{l+1}}a_\a^{-1} a_\b}\right.\nn
\\
&\left.\qquad \qquad\qquad \qquad\qquad \qquad\qquad
~~\, \times (-1) y^{\frac{\tilde m_\a+2}{l+1}-\frac{1}{2}}d^{-2}(a_\a b_\b^{-1})\frac{1-y^{\frac{l-1-\tilde m_\a}{l+1}}d^{2}a_\a^{-1}b_\b}{1-y^{\frac{\tilde m_\a+2}{l+1}}d^{-2}a_\a b_\b^{-1}}\right)\nn
\\
&\qquad \qquad\qquad \qquad\qquad \qquad\qquad
\times \prod_{j=0}^{l-1} (-1)y^{\frac{j}{l+1}-\frac{1}{2}}d^{2}(a_\a^{-1} b_\b)\frac{1-y^{\frac{l+1-j}{l+1}}d^{-2}a_\a b_\b^{-1}}{1-y^{\frac{j}{l+1}}d^{2}a_\a^{-1} b_\b}
\end{align}
It is simplified as
\begin{align}
Z^B(i\infty,z,\xi_\a,\eta_\b,\chi)
&=(-1)^ky^{2(lN-k)N\left(1-\frac{2}{l+1}\right)-\frac{2(lN-k)^2}{l+1}+N^2\sum_{j=0}^{l-1}\left(1-\frac{2j}{l+1}\right)}d^{2kN}
\\
&\sum_{\vec{\tilde n} \phantom{,} \textrm{s.t}\, |\vec{\tilde n}|=lN-k}  \prod_{\a,\b=1}^{N} \left(\prod_{\tilde m_\a =0}^{\tilde n_\a-1}
\frac{1-y^{\frac{l+1-\tilde n_\b + \tilde m_\a}{l+1}}a_\a a_\b^{-1}}{1-y^{\frac{\tilde n_\b-\tilde m_\a}{l+1}}a_\a^{-1} a_\b}
\frac{1-y^{\frac{l-1-\tilde m_\a}{l+1}}d^2a_\a^{-1} b_\b}{1-y^{\frac{\tilde m_\a+2}{l+1}}d^{-2}a_\a b_\b^{-1}}\right)\nn
\\
&\qquad \qquad\qquad \qquad \times
\prod_{j=0}^{l-1}\frac{1-y^{\frac{l+1-j}{l+1}}d^{-2}a_\a b_\b^{-1}}{1-y^{\frac{j}{l+1}}d^2a_\a^{-1} b_\b}
\nn
\end{align}
The chiral primary contributions are
\begin{align}\label{CR_U(k)NNAd_Btheory}
\CI^B_{CR}=
&\sum_{\vec{\tilde n} \phantom{,} \textrm{s.t}\, |\vec{\tilde n}|=lN-k}  \prod_{\a,\b=1}^{N} \left(\prod_{\tilde m_\a =0}^{\tilde n_\a-1}
\frac{1-(yq^{1/2})^{\frac{l+1-\tilde n_\b + \tilde m_\a}{l+1}}a_\a a_\b^{-1}}{1-(yq^{1/2})^{\frac{\tilde n_\b-\tilde m_\a}{l+1}}a_\a^{-1} a_\b}
\frac{1-(yq^{1/2})^{\frac{l-1-\tilde m_\a}{l+1}}d^2a_\a^{-1} b_\b}{1-(yq^{1/2})^{\frac{\tilde m_\a+2}{l+1}}d^{-2}a_\a b_\b^{-1}}\right) \nn
\\
&\qquad\qquad\qquad\qquad \times \prod_{j=0}^{l-1}\frac{1-(yq^{1/2})^{\frac{l+1-j}{l+1}}d^{-2}a_\a b_\b^{-1}}{1-(yq^{1/2})^{\frac{j}{l+1}}d^2a_\a^{-1} b_\b}
\end{align}
The factors $\frac{1-(yq^{1/2})^{\frac{l-1-\tilde m_\a}{l+1}}d^2a_\a^{-1} b_\b}{1-(yq^{1/2})^{\frac{\tilde m_\a+2}{l+1}}d^{-2}a_\a b_\b^{-1}}$ are contributions of operators $q_\a Y^{\tilde m_\a}\tilde q_\b$, $\mathcal Q_- (\wbar{q_\a Y^{\tilde m_\a}\tilde q_\b})$ and the factors
$\prod_{j=0}^{l-1}\frac{1-(yq^{1/2})^{\frac{l+1-j}{l+1}}d^{-2}a_\a b_\b^{-1}}{1-(yq^{1/2})^{\frac{j}{l+1}}d^2a_\a^{-1} b_\b}$ are contributions of singlets $M_j^{\a\b}$, $\mathcal Q_- (\wbar{M_j^{\a\b}})$. Because $\CI^A_{CR}=\CI^B_{CR}$ since this is the $\tau\rightarrow \infty$ limit
of $Z_A=Z_B$, the generators should match.
As explained below \eqref{EG_U(k)NfNaAd_dual} all contributions from $q_\a Y^{\tilde m_\a}\tilde q_\b$ are canceled out. When $k \geq l$ all singlets of the B theory $M_j$, $j=0,\ldots,l-1$ contribute to the index. However, when $k<l$ only $M_j$, $j=0,\ldots, k-1$ contribute to the index.

Let us compute the chiral primary contributions for some examples and define $x=yq^{1/2}$ for simplicity.
For $k=l N$ the B theories become non-gauge theory and consist of singlet fields, $M_j$, $j=0, \ldots, l-1$. Chiral primary contribution is given by
\begin{align}\label{ChiralPrimary:k=lN}
\CI_{CR}(q,y,a_\a,b_\b,d)= \prod_{\a,\b=1}^{N} \prod_{m =0}^{l-1}
\frac{1-x^{\frac{l+1-m}{l+1}}d^{-2}a_\a b_\b^{-1}}{1-x^{\frac{m}{l+1}}d^2a_\a^{-1} b_\b}
\end{align}
The denominators come from $M_j = Q X^{j} \tilde Q$ and the numerators come from $\mathcal Q_- (M_j)=\mathcal Q_- (Q X^{j} \tilde Q)$. The chiral primary operators do not have any non-trivial relation, which is manifest in the description of the B theory. Terms in the numerators of \eqref{ChiralPrimary:k=lN} cannot be interpreted as relations because any product of generators does not have corresponding charges, $x^{\frac{l+1-m}{l+1}}d^{-2}a_\a b_\b^{-1}$. The elliptic genus of the theory is given by products of contributions of singlet fields $M_j$, $j=0,\ldots, l-1$,
\begin{align}
Z(\tau, z, \xi_\a, \eta_\b,\chi)= \prod_{\a, \b=1}^{N}  \prod_{j=0}^{l-1}\frac{\theta_1(\tau| -\xi_\a+\eta_\b +2\chi+\frac{(j-l-1)z}{l+1})}{\theta_1(\tau|-\xi_\a+\eta_\b+2\chi+\frac{jz}{l+1})} ~.
\end{align}

Let us consider $N=1$ cases. B theory is $U(l-k)$ gauge theory so $k\leq l$ is the valid range of the duality. The elliptic genus and chiral primary contribution are given by
\begin{align}
&Z(\tau,z,\chi)= \prod_{m=0}^{k-1}\frac{\theta_1(\tau|\frac{k-m}{l+1}z-z)}{\theta_1(\tau|\frac{k-m}{l+1}z)}\frac{\theta_1(\tau|2\chi+\frac{m}{l+1}z-z)}{\theta_1(\tau|2\chi+\frac{m}{l+1}z)}\\
&\CI(x,d)=
\prod_{m=0}^{k-1}\frac{1-x^{1-\frac{k-m}{l+1}}}{1-x^{\frac{k-m}{l+1}}}
\frac{1-x^{\frac{l+1-m}{l+1}}d^{-2}}{1-x^{\frac{m}{l+1}}d^2}
\end{align}
where $d=e^{2\pi i \chi}$ is a fugacity for the global $U(1)$ symmetry under which the chiral multiplets have the same charge. The first factor is the contribution of $\Tr X^j$, $j=1,\ldots, \min(k, l-k)$. For $l-k<k$, contributions of $\Tr X^j$, $j=l-k+1,\ldots, k$ are canceled out. The second factor comes from the operators $M_j=QX^j\tilde Q$, $j=0,\ldots, k-1$. The elliptic genus is the product  of the elliptic genus of  chiral fields $u_i$, $i = 1, \ldots, \min(k,l-k)$ and $v_j$, $j=0,\ldots, k-1$ where the chiral fields are identified as $u_i = \Tr X^i$  and $v_j = M_j$.
The expression of $u_i$ is reminiscent of the elliptic genus of the minimal models. It would be interesting to find the underlying CFT for the above theory. The particularly simple case is $l=k$ case. In this case, the B theory is the theory of $k$ singlets $M_j$ with $j=0 \cdots k-1$, whose
central charge contribution is $(1-\frac{2j}{k+1})$ so that the central charge of the CFT is
$\frac{c}{3}=\sum_{j=0}^{k-1}(1-\frac{2j}{k+1})=2k-\frac{2k^2}{k+1}$. The chiral ring relation between different $M_i$ and $M_j$ is trivial
since $M_i M_j \sim M_n$ is impossible since each $M_i$ has  $U(1)_a$ charge 2.

Now turn into  $k=1$ case. The elliptic genus is given by
\begin{align}
Z(\tau,z,a_\a,b_\b,d)= - \frac{\theta_1(\tau|-\frac{l}{l+1}z)}{\theta_1(\tau|\frac{1}{l+1}z)}
\sum_{\a=1}^N  \prod_{\b=1, \b\neq\a}^{N}
\frac{\theta_1(\tau|\xi_\a -\xi_\b-z)}{\theta_1(\tau|\xi_\a -\xi_\b )}
\prod_{\gamma =1}^N
\frac{\theta_1(\tau|-\xi_\a +\eta_\b +2\chi-z)}{\theta_1(\tau|-\xi_\a +\eta_\b +2\chi)}
\end{align}
The first factor comes from the adjoint field and is the same as the elliptic genus of the $l$-th minimal model. Since the adjoint field is neutral,
the underlying CFT is indeed the minimal model \cite{Witten:1993jg}. In this case, CFT consists of the tensor product of the $l$-th minimal model and the CFT of
$N=(2,2) \,\, U(1)$ with
$N$ flavors. The central charge of $k=1$ case is indeed given by the sum of that of the $l$-th minimal model and that of  the $U(1)$ with
$N$ flavors. This theory has another dual description, which is $U(N-1)$ gauge theory with $N$ fundamental/anti-fundamental chiral fields, $q$, $\tilde q$ and decoupled chiral field $Y$ with superpotential $W=Y^{l+1}+Mq\tilde q$.

Let us work out the matching of the chiral ring elements for some simple dual pairs.
Let us consider A theory for $U(1)$ gauge with $N=1$ pair of fundamental and anti-fundamental chiral multiplets and an adjoint with superpotential $W=\Tr X^{4}$ i.e. $l=3$. B theory is a $U(2)$ gauge theory with additional gauge singlet matter fields $M_j$, $j=0,1,2$ with superpotential $W=\Tr Y^{4}+M_0 q Y^2 \tilde q + M_1\tilde q Y q+M_2\tilde q q$. Chiral primaries of the A theory are obtained as
\begin{align}
\CI_{CR}^A=\frac{1-x^{3/4}}{1-x^{1/4}}\frac{1-xd^{-2}}{1-d^2}
\end{align}
Because the gauge group is $U(1)$ the linearly independent bosonic chiral ring generators are $ X$ and $Q \tilde Q$. The first factor $\frac{1-x^{3/4}}{1-x^{1/4}}=1+x^{1/4}+x^{2/4}$ correspond to the identity, $X$ and $X^2$. It reflects the constraint $X^3=0$. Another factor $\frac{1-xd^{-2}}{1-d^2}$ comes from the mesonic operator $Q\tilde Q$ and $\mathcal Q_-(\wbar{Q\tilde Q})$.
Chiral primaries of the B theory are computed as
\begin{align}
\CI^B_{CR}=&\frac{1-x^{2/4}}{1-x^{2/4}}\frac{1-x^{3/4}}{1-x^{1/4}}\cdot\frac{1-x^{2/4}d^2}{1-x^{2/4}d^{-2}}\frac{1-x^{1/4}d^2}{1-x^{3/4}d^{-2}}\cdot \frac{1-x d^{-2}}{1-d^{2}}\frac{1-x^{3/4} d^{-2}}{1-x^{1/4}d^{2}}\frac{1-x^{2/4} d^{-2}}{1-x^{2/4}d^{2}}\nn\\
=&\frac{1-x^{3/4}}{1-x^{1/4}}\frac{1-xd^{-2}}{1-d^2}
\end{align}
At first an operator $\Tr  Y^2$ is not constrained by the superpotential or the $U(2)$ B theory characteristic equation which is
\begin{align}
Y^2 - Y \Tr Y + \frac{1}{2}\left( (\Tr X)^2 - \Tr X^2 \right) I_{2} = 0
\label{Characteristic_U(2)}
\end{align}
where $I_2$ is a $2\times 2$ identity matrix. However it should not be a chiral primary in the IR to be consistent with the duality. Actually $\frac{1-x^{2/4}}{1-x^{2/4}}$ is the contribution of $\Tr  Y^2$ and $\mathcal Q_-(\wbar{\Tr  Y^2})$, which cancel each other. This is consistent with the $U(1)$ A theory in which $\Tr  X^2$ is not a linearly independent operator. The index also shows pair cancellations, $(\tilde q q, M_2)$ as $\frac{1-x^{2/4}d^{2}}{1-x^{2/4} d^{-2}} \frac{1-x^{2/4} d^{-2}}{1-x^{2/4}d^{2}}=1$  and $(\tilde q Y q, M_1)$ as $\frac{1-x^{1/4}d^{2}}{1-x^{3/4} d^{-2}}\frac{1-x^{3/4} d^{-2}}{1-x^{1/4}d^{2}}=1$.  Due to the characteristic equation \eqref{Characteristic_U(2)} $\tilde q Y^2 q$ is not a linearly independent operator so it does not annihilate $M_0$. Thus $M_0$ operator survives and corresponds to $Q\tilde Q$ operator of A theory.

Let us consider A theory with $U(2)$ gauge group and $N=1$ pair of fundamental and anti-fundamental chiral multiplets and an adjoint with a superpotential  $W=\Tr X^{4}$ i.e. $l=3$. B theory is $U(1)$ gauge theory with additional gauge singlet matter fields $M_j$, $j=0,1,2$ with superpotential $W=\Tr Y^{4}+M_0 q Y^2 \tilde q + M_1\tilde q Y q+M_2\tilde q q$. Chiral primary contribution is given by
\begin{align}
\CI_{CR}^A&=\frac{1-x^{2/4}}{1-x^{2/4}}\frac{1-x^{3/4}}{1-x^{1/4}} \frac{1-xd^{-2}}{1-d^2}\frac{1-x^{3/4}d^{-2}}{1-x^{1/4}d^2}\nn\\
&=\frac{1-x^{3/4}}{1-x^{1/4}} \frac{1-xd^{-2}}{1-d^2}\frac{1-x^{3/4}d^{-2}}{1-x^{1/4}d^2}
\end{align}
The first factor $\frac{1-x^{3/4}}{1-x^{1/4}}=1+x^{1/4}+x^{2/4}$ correspond to the identity, $\Tr X$ and $(\Tr X)^2$. Even though $\Tr X^2$ is not constrained by the superpotential or the $U(2)$ characteristic equation its contribution $\frac{1-x^{2/4}}{1-x^{2/4}}$ is canceled out. It is a quantum constraint consistent with the $U(1)$ B theory. The second and third factors correspond to $Q\tilde Q$, $QX\tilde Q$, $\mathcal Q_-(\wbar{Q\tilde Q})$, $\mathcal Q_-(\wbar{QX\tilde Q})$.
Chiral primaries computed from the B theory are
\begin{align}
\CI^B_{CR}=&\frac{1-x^{3/4}}{1-x^{1/4}}\cdot\frac{1-x^{2/4}d^2}{1-x^{2/4}d^{-2}}\cdot \frac{1-x d^{-2}}{1-d^{2}}\frac{1-x^{3/4} d^{-2}}{1-x^{1/4}d^{2}}\frac{1-x^{2/4} d^{-2}}{1-x^{2/4}d^{2}}\nn\\
=&\frac{1-x^{3/4}}{1-x^{1/4}}\frac{1-xd^{-2}}{1-d^2}\frac{1-x^{3/4}d^{-2}}{1-x^{1/4}d^2}
\end{align}
where $\tilde q q$ and $M_2$ are canceled out as $\frac{1-x^{2/4}d^{2}}{1-x^{2/4} d^{-2}}\frac{1-x^{2/4} d^{-2}}{1-x^{2/4}d^{2}}=1$. Thus the chiral primaries are $\Tr  Y$, $M_0$, $M_1$  consistent with the duality.

Let us consider A theory with $U(2)$ gauge group and $N=2$ flavors and an adjoint with a superpotential $W=\Tr X^3$. B theory is also $U(2)$ theory so the characteristic equation is the same as A theory. The chiral primary contribution is given by
\begin{align}
\frac{1+x^{\frac{1}{3}}(1-d^4\frac{b_1b_2}{a_1a_2})+x^{\frac{2}{3}}(1-d^2\frac{b_1}{a_1}-d^2\frac{b_1}{a_2}-d^2\frac{b_2}{a_1}-d^2\frac{b_2}{a_2}-d^4\frac{b_1b_2}{a_1a_2})+O(x)}{\prod_{\a,\b=1}^{2}(1-d^2\frac{b_\b}{a_\a})(1-x^{\frac{1}{3}}d^2\frac{b_\b}{a_\a})}~.
\end{align}
The denominator comes from $(M_0)_{\a\b}=Q_\a \tilde Q_\b$ and $(M_1)_{\a\b}=Q_\a X \tilde Q_\b$. The numerator contains terms identified as $x^{\frac{1}{3}} \rightarrow \Tr X$, $x^{\frac{2}{3}} \rightarrow (\Tr X)^2$, $-x^{\frac{1}{3}}d^4\frac{b_1b_2}{a_1a_2} \rightarrow \det M_0 \Tr X + c \epsilon^{\a\b}\epsilon^{\g\d}(M_1)_{\a\g}(M_0)_{\b\d} =0 $ for some $c$, $-x^{\frac{2}{3}}d^2\frac{b_\b}{a_\a}\rightarrow (M_1)_{\a\b}\Tr X - (M_0)_{\a\b} (\Tr X)^2=0$. The rest of terms are the order of $x$ up to $x^3$.

\subsection{Summary of the Phase Structure}

We have considered theories with four parameters, $k$, $N_f$, $N_a$, $l$ where $U(k)$ gauge group with $N_f$ fundamental chiral fields and $N_a$ anti-fundamental chiral fields and the superpotential $W=\Tr X^{l+1}$. We assume twisted masses for the chiral fields are zero. The FI parameter $\xi$ runs for $N_f \neq N_a$ under the RG and does not for $N_f=N_a$. The theta angle $\theta$ is appropriately tuned to minimize the potential energy \cite{Witten:1993yc}.

 We just have to consider $N_f \geq N_a$ cases since $N_f < N_a$ cases reduce to those cases under charge conjugation. If $k> l N_f$ we expect the supersymmetry is spontaneously broken. If $k\leq l N_f$ we show that the theory has the dual description, $U(lN_f - k)$ gauge theory in the IR. If $N_a=0$ we argue that the theory has the mass gap and $\binom{l N_f}{k}$ isolated vacua. If $0<N_a < N_f$ we expect that the theory has not only isolated vacua but also non-trivial non-linear sigma model on the Higgs branch. If $N_a = N_f$ the theory is superconformal and the duality holds for $\xi\neq 0$ where there are no vacua on the Coulomb branch. See \cite{Aharony:2016jki}, for related discussion. This phase structure is summarized at Table1.
\begin{table}[h]
	\begin{center}
		\begin{tabular}{|c|c|c|}
			\hline
			$lN_f<k$ &  $N_f\geq N_a$ &spontaneous breaking of the supersymmetry\\
			\hline
			$lN_f=k$ &  $N_f\geq N_a>0 $ & theory of $N_f N_a$ free chiral fields \\
			\cline{2-3}
			&  $N_f>0,\, N_a=0 $ & one massive isolated vacua \\
			\hline
			$lN_f>k$ &  $N_f=N_a>0 $ & SCFT on Higgs branch, $\xi\neq 0$ \\
			\cline{2-3}
			&  $N_f>N_a>0 $ & massive isolated vacua and NLSM on Higgs branch \\
			\cline{2-3}
			&  $N_f>0,\, N_a=0 $ & $\binom{lN_f}{k}$ massive isolated vacua \\
			\hline
		\end{tabular}
	\end{center}\caption{Phase structure of $\mathcal N=(2,2)$ $U(k)$ gauge theory with $N_f$ fundamental, $N_a$ anti-fundamental and one adjoint chiral fields with the superpotential $W=\Tr X^{l+1}$}
\end{table}

\section{Relation to dualities in 3 and 4 dimensions}

In 3 and 4-dimensions, one can find the similar dualities studied in this paper. In 4-dimensions, one has Kutasov-Schwimmer-Seiberg dualities\cite{Kutasov:1995np}. In 3-dimensions, the analogue was worked out in \cite{Kim:2013cma}. The natural question is if the 2-dimensional dualities considered in the paper are related to
the dualities in 3-dimensions and 4-dimensions via dimensional reduction. In fact the 3-dimensional dualities of \cite{Kim:2013cma} can be
obtained from the Kutasov-Schwimmer-Seiberg dualities \cite{ Nii:2014jsa, Amariti:2014iza} following \cite{Aharony:2013dha}.
One can also obtain 2-dimensional $\mathcal N=(2,2)$ gauge theories from 3-dimensional $\mathcal N=2$ gauge theories \cite{Willett:strings2016}.
In this section, we briefly summarize the reduction of the 4-dimensional dualities to the 3-dimensional dualities. After that, we show that the
2-dimensional dualities follow from
the 3-dimensional dualities following \cite{Willett:strings2016}. Thus we make the relation explicit between dualities in 2,3 and 4-dimensions.
\footnote{It is known that Seiberg-type dualities in two dimensions follow from similar dualities in four dimensions. Reducing 4-dimensional $\mathcal N=1$ theories to 2-dimensional $(0,2)$ theories, some 2-dimensional Seiberg-type dualities have 4-dimensional origin \cite{Kutasov:2013ffl, Kutasov:2014hha, Gadde:2015wta}. One can
directly relate the 4d Kutasov-Schwimmer-Seiberg and 2-dimensional dualities of the paper.}

\subsection{From 4-dimensional dualities to 3-dimensional dualities}

4-dimensional Kutasov-Schwimmer dualities consider the following A and B theory and two theories flow to the same SCFT in the IR.

Theory A: $U(N_c)$ gauge theory with $N_f$ fundamental $Q^{\alpha}$ and anti-fundamental fields $\tilde{Q}^{\beta}$, one adjoint field $X$ and with the superpotential
\begin{equation}
W=\Tr X^{l+1}.
\end{equation}

Theory B: $U(lN_f-N_c)$ gauge theory with $N_f$ fundamental $q_{\alpha}$ and anti-fundamental fields $\tilde{q}_{\beta}$, one adjoint field $Y$, singlet fields $M_j$ with $j=0\cdots l-1$ and
with the superpotential
\begin{equation}
W=\Tr Y^{l+1}+\sum_{j=0}^{l-1}M_j\tilde q Y^{l-1-j} q.
\end{equation}

The 3-dimensional analogue was worked out in \cite{Kim:2013cma}, where some evidences are presented that the following A and B theory are equivalent in the IR.

Theory A: $U(N_c)$ gauge theory with $N_f$ fundamental $Q^{\alpha}$  and anti-fundamental fields $\tilde{Q}^{\beta}$, one adjoint field $X$ and with the superpotential
\begin{equation}
W=\Tr X^{l+1}.
\end{equation}

Theory B: $U(lN_f-N_c)$ gauge theory with $N_f$ fundamental $q_{\alpha}$ and anti-fundamental fields $\tilde{q}_{\beta}$, one adjoint field $Y$, singlet fields $M_j$, $v_{j, \pm}$ with $j=0 \cdots l-1$ and
with the superpotential
\begin{align}
W_B=\Tr Y^{l+1} + \sum_{j=0}^{l-1}M_j\tilde q Y^{l-1-j} q +
 \sum_{j=0}^{l-1}(v_{j,+} \tilde V_{l-1-j,-} +v_{j,+} \tilde V_{l-1-j,-})
\end{align}
where $\tilde V_{j, \pm}$ are the monopole operators of $U(lN_f-N_c)$ while the monopole operators of $U(N_c)$ are mapped to $ v_{j, \pm}$.

One can consider  3-dimensional reduction of 4-dimensional Seiberg dual theories on $\mathbb R^3\times S^1$ with the circle radius $r\rightarrow 0$. But this naive dimensional reduction of the 4-dimensional dual theories results in 3-dimensional theories which are not dual to each other. This can be understood as follows \cite{Aharony:2013dha}. The 4-dimensional Seiberg duality is the IR duality, which implies two theories $A$ and $B$ are identical at energies below their strong coupling scales, $E \ll \Lambda_A, \Lambda_B$. When a 4-dimensional theory is compactified on the circle the strong coupling scale is given by $ \Lambda^b = \exp(-4\pi /(r g^2_3))$ where $b$ is the one-loop beta function coefficient and $g^2_3$ is a
3-dimensional gauge coupling. If we take the limit $r\rightarrow 0$ with a fixed $g^2_3$ the strong coupling scale becomes $\Lambda \rightarrow 0$ for a asymptotic free theory. In this limit, the low-energy limit $E \ll \Lambda_A, \Lambda_B$ where the duality is valid becomes meaningless.

In order to obtain 3-dimensional dualities from 4-dimensional duailities, we keep $r$ fixed and look at energies  $E \ll \Lambda_A, \Lambda_B, 1/r$ where the effective low-energy dynamics is three dimensional and deduces the 3d dualities. The finite radius leads to a compact Coulomb branch and an additional Affleck-Harvey-Witten (AHW) superpotential. The 4-dimensional theories of interest on a circle has $2l$ unlifted Coulomb branch parameterized by $v_{i,\pm}$ in theory A and $\tilde v_{i,\pm}$ in theory B and the superpotential \cite{Nii:2014jsa},
\begin{align}
&W_A=\Tr X^{l+1} + \eta \sum_{j=0}^{l-1}v_{j,+}v_{l-1-j,-}
\\
&W_B=\Tr Y^{l+1} + \sum_{j=0}^{l-1}M_j\tilde q Y^{l-1-j} q +
\tilde \eta \sum_{j=0}^{l-1}\tilde v_{j,+} \tilde v_{l-1-j,-} \label{TheoryB_supot_deformed}
\end{align}
where $\eta \equiv \Lambda^b$. Due to the AHW superpotential ($\eta$ term), axial $U(1)$ flavor symmetry is explicitly broken. Such AHW superpotential can vanish in the theory deformed by a large real mass of a charged field so that the axial $U(1)$ symmetry is restored. The duality given in \cite{Kim:2013cma} can be obtained from the theory A with $N_f+2$ flavors and real masses,
\begin{align}
m_\alpha = (0,\cdots,0,m,-m),\qquad \tilde m_\beta = (0,\cdots,0,-m,m)
\end{align}
where $m_\alpha$ and $\tilde m_\beta$ are real masses for fundamental chiral fields $Q_\alpha$ and anti-fundamental chiral fields $\tilde Q_\beta$ respectively. If a charged field is integrated out the monopole operators at high-energy and low-energy are related as $v_{\rm high} = m_{\mathbb C}^{1/2} \cdot v_{\rm low}$ where $m_{\mathbb C}$ is a complex mass of the charged field. The flavors have no complex mass but are integrated out due to the real masses so the AHW superpotential vanishes at low energies since $m_{\mathbb C}=0$ \cite{Aharony:2013dha}. Thus the theory A becomes $U(N_c)$ gauge theory with $N_f$ flavors and an adjoint with the superpotential, $W_A = \Tr X^{l+1}$, which is nothing but the theory A of \cite{Kim:2013cma}.

The theory B is $U(l(N_f+2)-N_c)$ gauge theory with $N_f+2$ flavors and an adjoint with the superpotential \eqref{TheoryB_supot_deformed} in the UV. In \cite{Nii:2014jsa} this theory is perturbed by $W=\sum_{j=0}^{l}\frac{s_j}{k+1-j}\Tr Y^{k+1-j}$. The effect of the perturbation is to break the gauge group $U(n_1)\times \cdots \times U(n_l)$, $ \sum_i n_i=l(N_f+2)-N_c $ and the adjoint field becomes massive. Furthermore, the real mass deformation leads to a vacuum expectation value of scalar fields in each $U(n_i)$ vector multiplet, $ \sigma_{U(n_i)}={\rm diag}(0,\ldots, 0,-m,m) $. This breaks the $U(n_i)$ gauge groups to $ U(n_i-2)\times U(1)_{i,1} \times U(1)_{i,2} $.
 $U(n_i-2)$ gauge group has $N_f$ flavors and  each of $U(1)_{i,a}$ with $ a=1,2$ has one flavor. Thus the $U(1)_{i,a}$ sector can be dualized to the $XYZ$ model. Under this duality, the monopole operators $v_{U(1)_{i,a},\pm}$ of $U(1)_{i,a}$ theory are mapped to singlet fields of $XYZ$ model. These singlet fields interact with the monopole operators of $U(n_i)$ through a superpotential. Thus we have $U(n_1-2)\times \cdots \times U(n_l-2)$ gauge theory interacting with $2l$ $XYZ$ models. It is argued in \cite{Nii:2014jsa}, by turning off the perturbation $s_j \rightarrow 0$, $(j\neq 0)$ the gauge group is enhanced to $U(lN_f-N_c)$ theory with $N_f$ flavors, one adjoint and  monopole operators, $\tilde V_{j,\pm}$. $2l$ singlet fields $v_{U(1)_{i,1},+}$, $v_{U(1)_{i,2},-}$ are identified with the singlet fields $v_{j,\pm}$ interacting with the monopole operators of $U(lN_f-N_c)$ gauge sector by the superpotential,
\begin{align}
W_B=\Tr Y^{l+1} + \sum_{j=0}^{l-1}M_j\tilde q Y^{l-1-j} q +
 \sum_{j=0}^{l-1}(v_{j,+} \tilde V_{l-1-j,-} +v_{j,+} \tilde V_{l-1-j,-})
\end{align}
which is the theory B in \cite{Kim:2013cma}. Hence one can recover the duality of \cite{Kim:2013cma}.

\subsection{From the 3-dimensional dualities  to the 2-dimensional dualities}

Here we review the work \cite{Willett:strings2016} and apply it to the theory with one adjoint field. Let's consider 3-dimensional $\mathcal N=2$ $U(N_c)$ gauge theory with $N_f$ pairs of fundamental and anti-fundamental chiral fields, $(Q^\alpha, \tilde Q^\beta)$. This theory has the IR dual description, $U(N_f-N_c)$ gauge theory with $N_f$ pairs of fundamental and anti-fundamental chiral fields $(q_\alpha, \tilde q_\beta)$, $N_f^2$ singlet chiral fields, $M^{\alpha \beta}$ and two singlet chiral fields $v_\pm$ which correspond to bare monopole fields in the original $U(N_c)$ gauge theory. If the theories are placed on $R^2\times S^1_r$ with radius $r$, one finds effective 2-dimensional descriptions at energies below $1/r$. However, reducing the theories on the Coulomb branch (3d FI parameter $\zeta =0$) is not consistent with the duality  \cite{Willett:strings2016}. The Coulomb branch can be lifted by turning on non-zero, finite FI parameter, which can be seen from the potential,
\begin{align}
V=&\frac{1}{2e^2}\Tr\,[\sigma,\bar \sigma]^2 + \frac{e^2}{2}\sum_{i,j=1}^{N_c}\left|\sum_{\alpha=1}^{N_f}Q^\alpha_i \bar Q^{\alpha j} -\sum_{\beta=1}^{N_f} \bar{\tilde{Q}}^{\beta j}\tilde{Q}^\beta_i -\zeta\delta^i_j\right|^2+\frac{1}{2}\sum_{\alpha=1}^{N_f}\bar Q^\alpha\left\{\bar \sigma,\sigma\right\}Q^\alpha + \frac{1}{2}\sum_{\beta=1}^{N_f}\tilde{Q}^\beta\left\{\bar \sigma,\sigma\right\}\bar{\tilde{Q}}^\beta\nonumber
\\
=& \frac{e^2}{2}\sum_{i,j=1}^{N_c}\left|\sum_{\alpha=1}^{N_f}Q^\alpha_i\bar Q^{\alpha j} -\sum_{\beta=1}^{N_f}\bar {\tilde{Q}}^{\beta j}\tilde{Q}^\beta_i -\zeta\delta^i_j\right|^2
 +\sum_{i=1}^{N_c}\sum_{\alpha=1}^{N_f}|\sigma^i|^2(|Q^\alpha_i|^2 +|\tilde Q^\alpha_i|^2)
\end{align}
where the second line is obtained by diagonalized $\sigma$ due to the potential, $\Tr\,[\sigma,\bar \sigma]^2$. If the FI parameter $\zeta$ is non-zero, $\sigma$ should vanish, i.e. the Coulomb branch is lifted.
Now we reduce the 3-dimensional theories to the 2-dimensional theories. To be more specific, the 3-dimensional $U(N_c)$ gauge theory is reduced to 2-dimensional $N=(2,2)$ $U(N_c)$ gauge theory with $N_f$ pairs of fundamental and anti-fundamental chiral fields and with non-zero 2-dimensional FI parameter $t=\zeta r$. On the other hand, the 3-dimensional $U(N_f-N_c)$ gauge theory has a non-zero FI parameter, which induces a non-zero real mass for the singlet chiral fields $v_+$, $v_-$. The FI parameter can be understood as a vacuum expectation value(vev) of the real scalar field in a background vector multiplet for the topological symmetry, $U(1)_J$. Monopole operators are charged under the topological symmetry. Hence $v_\pm$ picks up a mass term in the presence of nonzero FI parameter,
\begin{align}
\mathcal S=\int dx^3 d^2\theta d^2 \overline \theta \overline v_\pm e^{\pm i\zeta \theta \overline \theta} v_\pm.
\end{align}
Therefore, the singlet chiral fields $v_\pm$ get massive and are integrated out at energies below $1/r$ where $\zeta > 1/r$. The resulting 2-dimensional $\mathcal N=(2,2)$ theory is $U(N_f-N_c)$ with $N_f$ pairs of fundamental and anti-fundamental chiral fields $(q_\alpha, \tilde q_\beta)$ and the singlet chiral fields $M^{\alpha \beta}$, which is the dual description of the 2d $U(N_c)$ gauge theory.

Now let's consider the 3-dimensional theory with one adjoint chiral field $X$ and the superpotential, $W = \Tr X^{l+1}$ in addition to $N_f$ pairs of fundamental and anti-fundamental chiral fields $(Q^\alpha, \tilde Q^\beta)$. The dual description \cite{Kim:2013cma} is the $U(l N_f - N_c)$ gauge theory with $N_f$ pairs of fundamental and anti-fundamental chiral fields $(q_\alpha, \tilde q_\beta)$, $l N_f^2$ singlet chiral fields, $M_j^{\alpha \beta}$, $2l$ singlet chiral fields $v_{j,+}$, $ v_{j,-} $ where $j=0, \ldots, l-1$ and an adjoint field $Y$ and superpotential, $W=\Tr Y^{l+1} + \sum_j (M_{j}q Y^{l-1-j}\tilde q + v_{j,\pm} \tilde V_{l-1-j,\mp})$ where $ \tilde V_{j,\pm} $ are monopole fields in the dual theory and all flavor and gauge indices are contracted. All $v_{j,\pm}$ are charged under the topological symmetry with charges $\pm 1$, so as before, all of them are integrated out at energies below $1/r$ where $\zeta > 1/r$. Then the effective 2-dimensional description is $U(l N_f - N_c)$ gauge theory with $N_f$ pairs of fundamental and anti-fundamental chiral fields $(q_\alpha, \tilde q_\beta)$, $l N_f^2$ singlet chiral fields, $M_j^{\alpha \beta}$ and superpotential,
\begin{equation}
 W=\Tr Y^{l+1} +\sum_j M_{j}q Y^{l-1-j}\tilde q.
  \end{equation}
Therefore, the 2d dualities with $N_f=N_a$ can be derived from the 3d dualities. Other dualities with $N_f \neq N_a$ can be obtained by giving twisted masses to some chiral fields by weakly gauging corresponding $U(1)$ parts of $SU(N_f)\times SU(N_a)$.

\section{Mass gap }

In this section, we argue the existence of the mass gap for the $U(k)$ theory with one adjoint and $N_f > k$ fundamental chiral multiplets.
We closely follow the argument in \cite{Witten:1993xi}.
Let us consider 2d $\CN=(2,2)$ $U(k)$ gauge theory with $N_f$ fundamentals $Q_i$ and one adjoint $X$ and a superpotential,\footnote{This computation is also valid for the special case $W=\Tr X^{l+1}$.}
\begin{align}
W=\sum_{j=0}^l \frac{s_j}{l+1-j}\Tr X^{l+1-j}
\end{align}
For generic coefficients $\{s_j\}$ the superpotential can have $l$ distinct minima as
\begin{align}
\frac{\p W}{\p X}=\sum_{j=0}^{l}s_j X^{l-j} \equiv s_0\prod_{j=1}^l(X-\lambda_j \bold{I}_{k\times k})~.
\end{align}
The classical potential is given by

\begin{align}
V&=\frac{1}{2}\Tr [\s,\wbar \s]^2 + \frac{1}{2}\sum_{i,j=1}^{k}\,\left(\sum_{\a=1}^{N}\phi^{i}_\a \wbar\phi_{j}^\a + [X, \wbar X]^i_j-r\delta^i_j\right)^2 \label{potential_adjoint_2}
\\
&\qquad
+\frac{1}{2}\sum_{\a=1}^{N}\wbar \phi_{i}^\a \left\{\wbar \s ,\s \right\}^i_j\phi^{j}_\a
+ \frac{1}{2}\Tr ([\wbar X,\wbar \sigma] )([ \sigma, X])
+ \frac{1}{2}\Tr ( [\wbar X, \sigma]) ([\wbar \sigma, X]) \nn
\\
&\qquad
+\left|s_0\prod_{j=1}^l(X-\lambda_j \bold{I}_{k\times k})\right|^2\nn
\end{align}

Classically to have the zero potential we need $X=\lambda_j i_{k \times k}$for some $j$ so that
\begin{equation}
\sum_{\a=1}^{N}\phi^{i}_\a \wbar\phi_{j}^\a -r\delta^i_j=0
\end{equation}
Thus some of $\phi^{i}_\a$ should be nonzero.
Quantum mechanically this is not possible since this implies the breaking of the global
symmetry of $U(N_f)$, which is forbidden in 2 dimensions.
The resolution of this puzzle is standard. Even though $\phi^{i}_\a$ has zero expectation value
the bilinear $O^i_j =\phi^{i}_\a \wbar\phi_{j}^\a$ can have the nontrivial expectation value.
Following \cite{Witten:1993xi}, we have
\begin{equation}
<O>=N_f\int \frac{d^2 k}{(2\pi)^2}\frac{1}{k^2+\{\sigma, \bar{\sigma}\}}
\end{equation}
from \eqref{potential_adjoint_2}.
Using the same regularization scheme of \cite{Witten:1993xi},
we have
\begin{equation}
<O>=-\frac{N_f}{4\pi} \ln (\{\sigma, \bar{\sigma}\}/2\mu^2).
\end{equation}
The condition for vanishing energy in this approximation is
\begin{equation}
\{\sigma, \bar{\sigma}\}=2\mu^2 \exp (-4\pi r/N_f)
\end{equation}
From the classical Lagrangian we have $[\sigma, \bar{\sigma}]=0$ and $X=\lambda_j i_{k\times k}$
so that $\sigma = \mu \exp (-2\pi r/N_f) g$ with $g$ being a unitary matrix.
The nonzero vev of $O$ gives rise to masses for the fundamental flavors.
The above computation is a typical large $N_f$ computation. But the result is valid for finite $N_f$ by going to a sufficiently large negative value of $r$ \cite{Witten:1993xi}.

\appendix
\section{$\chi_y$ genus}
In the appendix we compute $\chi_y$ genus of the $U(k)$ theory with one adjoint and $N_f > k$ fundamental chiral multiplets.
We call it A theory. The contributions of the ground states of any $\CN=(2,2)$ theory can be obtained by
\begin{align}
Z(q,y)=Z(q\rightarrow 0,y)\equiv \chi_y~.
\end{align}
In the limit $q\rightarrow 0$, equivalently, $\tau \rightarrow i\infty$, the elliptic genus of a free chiral multiplet is reduced to
\begin{align}
\lim_{\tau \rightarrow i\infty} \frac{\theta_1(\tau|\xi_1)}{\theta_1(\tau|\xi_2)}=\frac{x_1^{1/2}-x_1^{-1/2}}{x_2^{1/2}-x_2^{-1/2}}
\end{align}
where $x_i=e^{2i\pi \xi_i}$. The elliptic genus of A theory is reduced to $\chi_y$ genus
\begin{align}
&  Z^A_{l,k,N_f}(q=0,y)\nn\\
&=(-1)^k\sum_{\vec{n} \phantom{,} \textrm{s.t}\, |\vec{n}|=k} \prod_{\a,\b=1}^{N_f} \prod_{m_\a =0}^{n_\a-1}
\frac{a_\a^{1/2}a_\b^{-1/2}y^{(n_\b -m_\a-l-1)/2(l+1)}-a_\a^{-1/2}a_\b^{1/2}y^{-(n_\b -m_\a-l-1)/2(l+1)}}{a_\a^{1/2}a_\b^{-1/2}y^{(n_\b -m_\a)/2(l+1)}-a_\a^{-1/2}a_\b^{1/2}y^{-(n_\b -m_\a)/2(l+1)}}\nonumber\\
&=(-1)^k\sum_{\vec{n} \phantom{,} \textrm{s.t}\, |\vec{n}|=k} \prod_{\a,\b=1}^{N_f} \prod_{m_\a =0}^{n_\a-1}
\frac{a_\a y^{-1/2-m_\a/(l+1)}-a_\b y^{1/2-n_\b/(l+1)}}{a_\a y^{-m_\a/(l+1)}-a_\b y^{-n_\b/(l+1)}}
\end{align}

As in $CP^{N-1}$ and Grassmannian models the ground state contributions are independent of the global symmetry fugacities $a_\a$.  As a trick to obtain the the explicit expressions, we take limits $a_\a\rightarrow \infty$ and $a_\a\rightarrow 0$ to obtain $\chi_y$ genus. We would like to consider $N_f+1$ flavors case and take limits on $a_{N_f+1}$. It depends on values of $n_{N_f+1}$,
\begin{align}
  Z^A_{l,k,N_f}(0,y)
&=(-1)^k\sum_{n_{N_f+1}=0}^k \left[\sum_{\vec{n} \phantom{,} \textrm{s.t}\, |\vec{n}|=k-n_{N_f+1}} \left(\prod_{\a,\b=1}^{N_f} \prod_{m_\a =0}^{n_\a-1}
\frac{a_\a y^{-1/2-m_\a/(l+1)}-a_\b y^{1/2-n_\b/(l+1)}}{a_\a y^{-m_\a/(l+1)}-a_\b y^{-n_\b/(l+1)}}\right)\right.\nn
\\
&\qquad\qquad \qquad \qquad \qquad
\left(\prod_{\a=1}^{N_f} \prod_{m_\a =0}^{n_\a-1}
\frac{a_\a y^{-1/2-m_\a/(l+1)}-a_{N_f+1} y^{1/2-n_{N_f+1}/(l+1)}}{a_\a y^{-m_\a/(l+1)}-a_{N_f+1} y^{-n_{N_f+1}/(l+1)}}\right)\nn
\\
&\qquad\qquad \qquad \qquad \qquad
\left(\prod_{\b=1}^{N_f} \prod_{m_{N_f+1} =0}^{n_{N_f}-1}
\frac{a_{N_f+1} y^{-1/2-m_{N_f+1}/(l+1)}-a_\b y^{1/2-n_\b/(l+1)}}{a_{N_f+1} y^{-m_{N_f+1}/(l+1)}-a_\b y^{-n_\b/(l+1)}}\right)\nn
\\
&\qquad\qquad \qquad \qquad \qquad
\left. \left(\prod_{m_{N_f+1} =0}^{n_{N_f+1}-1}
\frac{a_{N_f+1} y^{-1/2-m_{N_f+1}/(l+1)}-a_{N_f+1} y^{1/2-n_{N_f+1}/(l+1)}}{a_{N_f+1}y^{-m_{N_f+1}/(l+1)}-a_{N_f+1}y^{-n_{N_f+1}/(l+1)}}\right)\right]~.\label{eq:chi_y_genus_1}
\end{align}
The fourth line can be written in terms of q-binomial,
\begin{align}
{l \choose n}_y=\prod_{m=0}^{n-1}\frac{y^{(l-m)/2}-y^{-(l-m)/2}}{y^{(n-m)/2}-y^{-(n-m)/2}}~.
\end{align}
It can be considered as contributions of $\Tr X, \cdots, \Tr  X^n$ where $n = \textrm{min}(l-1, k,l-k)$.
Note that $\binom{l}{n}_y=\binom{l}{l-n}_y$.
We take limit $a_{N_f+1}\rightarrow \infty$ or $a_{N_f+1}\rightarrow 0$
\begin{align}
&\lim_{a_{N_f+1}\rightarrow \infty} Z^A_{l,k,N_f+1}(0,y)=\sum_{n=0}^k Z^A_{l,k-n,N_f}(0,y)  \cdot y^{(k-n)/2} \cdot  y^{-nN_f /2} \cdot {l \choose n}_{y^{\frac{1}{l+1}}}\label{eq:chi_y_genus_2}\\
&\lim_{a_{N_f+1}\rightarrow 0} Z^A_{l,k,N_f+1}(0,y)=\sum_{n=0}^k Z^A_{l,k-n,N_f}(0,y)  \cdot y^{-(k-n)/2} \cdot  y^{nN_f /2} \cdot {l \choose n}_{y^{\frac{1}{l+1}}}\label{eq:chi_y_genus_3}
\end{align}
Because $Z_{l,k,N_f}^A(0,y)$ is independent of $a_\a$,
\begin{align}
\lim_{a_{N_f+1}\rightarrow \infty} Z^A_{l,k,N_f+1}(0,y)=\lim_{a_{N_f+1}\rightarrow 0} Z^A_{l,k,N_f+1}(0,y)~.
\end{align}
Equating \eqref{eq:chi_y_genus_2} and \eqref{eq:chi_y_genus_3} gives
\begin{align}
Z^A_{l,k,N_f}(0,y)=\sum_{m=0}^{k-1} Z^A_{l,m,N_f}(0,y)\frac{[(k-m)N_f-m]_y}{[k]_y} {l \choose k-m}_{y^{\frac{1}{l+1}}}
\end{align}
where the summation is rearranged by $m=k-n$ and $[k]_y$ is the $q$-number,
\begin{align}
[k]_y=\frac{y^{k/2}-y^{-k/2}}{y^{1/2}-y^{-1/2}}
\end{align}
Thus $\chi_y$ genus of $U(k)$ gauge theory can be written as
\begin{align}
Z^A_{l,k,N_f}(0,y)&=\sum_{m_1=0}^{m_0-1}\frac{[(m_0-m_{1})N_f-m_{1}]_y}{[m_0]_y}  {l \choose m_0-m_{1}}_{y^{\frac{1}{l+1}}}
\\
&\qquad\qquad\quad\sum_{m_2=0}^{m_1-1}\frac{[(m_1-m_{2})N_f-m_{2}]_y}{[m_1]_y}  {l \choose m_1-m_{2}}_{y^{\frac{1}{l+1}}}\nn
\\
&\qquad\qquad\qquad \cdots \sum_{m_k=0}^{m_{k-1}-1}\frac{[(m_{k-1}-m_{k})N_f-m_{k}]_y}{[m_{k-1}]_y}  {l \choose m_{k-1}-m_{k}}_{y^{\frac{1}{l+1}}}\nn
\end{align}
where we define $m_0=k$ and $m_{j+1}$ summation exist only when $m_j\neq 0$. We would like to write down explicitly $\chi_y$ genus for some $k$. For a trivial theory, $k=0$ we set $Z^A_{l,0,N_f}(0,y)=1$. When A theory is a $U(1)$ gauge theory,
\begin{align}
Z^A_{l,1,N_f}(0,y)=\frac{y^{N_f/2}-y^{-N_f/2}}{y^{1/2}-y^{-1/2}} \cdot \frac{y^{l/2(l+1)}-y^{-l/2(l+1)}}{y^{1/2(l+1)}-y^{-1/2(l+1)}}
\end{align}
where the first factor is the $\chi_y$ genus of $CP^{N_f-1}$ model and the other factor comes from the adjoint field $X$. $\chi_y$ genus of $U(2)$ gauge theories are
\begin{align}
Z^A_{l,2,N_f}(0,y)&=\frac{[2N_f]_y}{[2]_y}{l \choose 2}_{y^\frac{1}{l+1}}+\frac{[N_f]_y[N_f-1]_y}{[2]_y[1]_y}{l \choose 1}_{y^\frac{1}{l+1}}{l \choose 1}_{y^\frac{1}{l+1}}
\\
&=\frac{y^{N_f}-y^{-N_f}}{y^{1}-y^{-1}} \frac{(y^{\frac{l}{2(l+1)}}-y^{-\frac{l}{2(l+1)}})(y^{\frac{l-1}{2(l+1)}}-y^{-\frac{l-1}{2(l+1)}})}{(y^{\frac{1}{l+1}}-y^{-\frac{1}{l+1}})(y^{\frac{1}{2(l+1)}}-y^{-\frac{1}{2(l+1)}})}\nonumber
\\
&\qquad+\frac{(y^{\frac{N_f}{2}}-y^{-\frac{N_f}{2}})(y^{\frac{N_f-1}{2}}-y^{-\frac{N_f-1}{2}})}{(y^{1}-y^{-1})(y^{\frac{1}{2}}-y^{-\frac{1}{2}})}
\left( \frac{y^{\frac{l}{2(l+1)}}-y^{-\frac{l}{2(l+1)}}}{y^{\frac{1}{2(l+1)}}-y^{-\frac{1}{2(l+1)}}}\right)^2   \nn
\end{align}


\section*{Acknowledgements}

We thank for Seok Kim, Hee-Cheol Kim, Chiung Hwang, Francesco Benini and Cumrun Vafa for helpful discussions.
JP and HK is supported in part
by the NRF Grant 2015R1A2A2A01007058.
The work of HK is supported in part by the NRF-2013-Fostering Core Leaders of the Future Basic Science Program and by the Center for Mathematical Sciences and Applications at Harvard University. The work of KC is supported by the NRF Grant 2015K1A3A1A21000302 and 2016R1D1A1B0101519.

\end{document}